\documentclass[11pt]{combine}
\usepackage{amsmath,amsfonts,amssymb,amsthm,epsfig, graphicx, hyperref}
\usepackage[dvipsnames,table,dvipsnames*, svgnames*, hyperref]{xcolor}
\usepackage{natbib,amsmath,amssymb,amsthm,graphicx,setspace,paralist,booktabs,rotating,subcaption,float,color}
\usepackage{multirow}
\usepackage{fancyhdr}
\usepackage{bibunits}
\usepackage[small]{titlesec}

\newtheorem{theorem}{Theorem}
\newtheorem{corollary}{Corollary}
\newtheorem{proposition}{Proposition}

\newtheorem{lemma}{Lemma}
{
	\theoremstyle{definition}
	\newtheorem{definition}{Definition}
	\newtheorem{example}{Example}
	\newtheorem{remark}{Remark}
}
\newcommand{\beq}{\begin{equation}}
\newcommand{\eeq}{\end{equation}}
\newcommand{\beas}{\begin{align*}}
\newcommand{\eeas}{\end{align*}}
\newcommand{\bea}{\begin{align}}
\newcommand{\eea}{\end{align}}
\newcommand{\bei}{\begin{itemize}}
	\newcommand{\eei}{\end{itemize}}
\newcommand{\ben}{\begin{enumerate}}
	\newcommand{\een}{\end{enumerate}}
\newcommand{\bet}{\begin{theorem}}
	\newcommand{\eet}{\end{theorem}}
\newcommand{\bel}{\begin{lemma}}
	\newcommand{\eel}{\end{lemma}}
\newcommand{\bep}{\begin{proposition}}
	\newcommand{\eep}{\end{proposition}}
\newcommand{\bed}{\begin{definition}}
	\newcommand{\eed}{\end{definition}}
\newcommand{\bec}{\begin{corollary}}
	\newcommand{\eec}{\end{corollary}}
\newcommand{\bex}{\begin{example}}
	\newcommand{\eex}{\end{example}}

\newcommand{\R}{\mathbb{R}}
\newcommand{\E}{\mathbb{E}}

\def\Var{{\rm Var}}

\begin{document}


\title{\scshape The Asymptotic Distribution of Modularity in Weighted Signed Networks}
\author{Rong Ma and Ian Barnett \\
Department of Biostatistics, Epidemiology and Informatics\\
University of Pennsylvania\\
Philadelphia, PA 19104}
\date{}

\maketitle
\thispagestyle{empty}

	\begin{abstract}
Modularity is a popular metric for quantifying the degree of community structure within a network. The distribution of the largest eigenvalue of a network's edge weight or adjacency matrix is well studied and is frequently used as a substitute for modularity when performing statistical inference. However, we show that the largest eigenvalue and modularity are asymptotically uncorrelated, which suggests the need for inference directly on modularity itself when the network size is large. To this end, we derive the asymptotic distributions of modularity in the case where the network's edge weight matrix belongs to the Gaussian Orthogonal Ensemble, and study the statistical power of the corresponding test for community structure under some alternative model. We empirically explore universality extensions of the limiting distribution and demonstrate the accuracy of these asymptotic distributions through type I error simulations. We also compare the empirical powers of the modularity based tests with some existing methods. Our method is then used to test for the presence of community structure in two real data applications.
	\bigskip
	
\end{abstract}

\section{Introduction}
Many scientific and social systems are composed of large numbers of interacting elements. These systems can be conceptualized as networks where nodes represent elements in the system and network edges represent interactions between elements. Networks appear consistently across scientific domains, ranging from protein interaction networks within a living cell to social networks of people communicating within society \citep{wasserman1994social,boccaletti2006complex}. Networks frequently divide into communities, or groups of nodes that cluster together. Detecting these network communities is a well-studied problem, with the most popular methods revolving around the maximization of a function known as modularity over all possible partitions of the network into communities \citep{newman2004finding,newman2006finding,newman2006modularity,good2010performance,chen2014community}. For most moderately large networks, enumerating all possible divisions into communities to find the maximum is not feasible. Many methods have been developed that aim to find optimal or near-optimal solutions with low computational complexity \citep{agarwal2008modularity,lancichinetti2009community}.

One of the most well-known approaches for identifying network community structure is the spectral approach proposed by \citet{newman2006finding,newman2006modularity}. If we consider an undirected random graph $G(E,V)$ where $|V|=n$, whose signed edge weight matrix is $W_G$, we define its modularity using the Newman-Girvan definition as:
\begin{equation}\label{eqmod}
Q(G) =  \mbox{sgn}(u_1^\top) W_G \mbox{sgn}(u_1)
\end{equation}
where $u_1\in \R^n$ is the eigenvector corresponding to the largest eigenvalue, $\lambda_1(W_G)$, of $W_G$ and $\text{sgn}(u_1)\in \{0,\pm1\}^n$ is the vector of signs of $u_1$. Since by definition $Q(G)$ only depends on the weight matrix $W_G$, throughout the paper we will not distinguish $Q(G)$ and $Q(W_G)$. In this setting, a common choice of null model considers $W$ to be a Wigner matrix. In addition, the treatment of networks with signed weights is distinct from networks with positive weights with respect to modularity. For example, \cite{traag2009community} considered community detection in complex networks with positive and negative links using a generalized Potts model.

Although modularity is frequently used, interpreting modularity tends to be a subjective exercise, most frequently done without the aid of statistical inference. In cases where inference is performed, simulation from some assumed null distribution is required \citep{dwyer2014large,rizkallah2016brain,telesford2016detection,lichoti2016social,springer2017dynamic,zhang2016hypothesis}. For very large networks, simulation of the null distribution is not computationally feasible. Because of this, some have looked towards analytical and asymptotic inference solutions based on the spectral decomposition:
\begin{equation}\label{eqmodspec}
Q(G) = \sum_{i=1}^n \lambda_i(W_G) \{\mbox{sgn}(u_1^\top)u_i\}^2,
\end{equation}
where $u_i\in \R^n$ is eigenvector corresponding to the $i$th largest eigenvalue $\lambda_i(W_G)$.
Because $\lambda_1$ has a disproportionately large role in $Q(G)$ and because $\lambda_1(W_G)$ frequently can be well-modeled by a Tracy-Widom distribution \citep{tracy1994level} for general Wigner matrices \citep{tao2011random}, there are some methods that use $\lambda_1$ as a proxy for $Q(G)$ when performing inference \citep{bickel2016hypothesis,lei2016goodness}. While approximating modularity with $\lambda_1$ is a tempting alternative, as we will show the smaller terms in equation \eqref{eqmodspec} play a nontrivial role in the null distribution of modularity and should not be ignored.

In this paper, we derive the asymptotic distribution of modularity defined in \eqref{eqmod} as $n \rightarrow \infty$ under the Gaussian Orthogonal Ensemble (GOE) random matrices.  Weighted networks with signed edges such as correlation networks can be well-modeled by GOE random matrices under a variety of null models. Correlation networks frequently appear in many contexts such stock market price networks \citep{chi2010network}, brain activity networks based on functional magnetic resonance imaging (fMRI) \citep{bullmore2009complex}, and gene expression networks \citep{langfelder2008wgcna}, to name a few. We demonstrate the convergence rate and accuracy of this distribution through simulations, and also analytically and numerically explore  the statistical power of the associated tests under some alternatives. In addition, we perform tests for modularity in two data examples: a U.S. congressional voting network and a morphological network of the human cranium.

Throughout the paper, we denote $S^n=\{(x_1,...,x_{n+1})\in \R^{n+1}: x_1^2+x_2^2+...+x_{n+1}^2=1\}$, and $O(n)$ as the orthogonal group, consisting of all the $n\times n$ orthogonal matrices. For a symmetric matrix $W\in \R^{n\times n}$, we denote $\lambda_1(W)\ge \lambda_2(W)\ge ...\ge \lambda_n(W)$ as its ordered eigenvalues. The function $\text{sgn}(\cdot)$ returns the sign of an object (scalar, vector or matrix). We denote $\to_d$ as convergence in distribution and $\to$ as a.s. convergence. For any vector $x=(x_1,...,x_n)$, we denote its $\ell_1$ norm as $\|x\|_1=\sum_{i=1}^n|x_i|$, denote its $\ell_2$ norm as $\|x\|_2=(\sum_{i=1}^nx^2_i)^{1/2}$.

\section{The Asymptotic Distribution of Network Modularity}

\subsection{The Limit Distribution under GOE Setting}  \label{classicGOE:sec}

We first study the asymptotic distribution of $Q(W)$ under the GOE setting, where $W$ is a standard Wigner matrix representing signed edge weights, whose upper off-diagonal entries and the diagonal entries are jointly independent with $W_{ij} \sim N(0,1)$ for $i>j$ and $W_{ij} \sim N(0,2)$ for $i=j$.
Our first main result concerns the limiting distribution of the modularity $Q(W)$.

\bet \label{GOE.thm}
For a random sample $W \in \R^{n\times n}$ from the GOE, let $Q = \textup{sgn}(u_{1}^\top)W \textup{ sgn}(u_{1})$, where $u_1\in \R^n$ is the first eigenvector of $W$. 
Then, for all $x\in \R$, we have
\beq \label{lim.law}
pr\big\{n^{-1}(Q-2{n}^{1/2}\|u_1\|_1^2)\le x\big\} \to \Phi\bigg\{\frac{x}{{2}^{1/2}(1-2/\pi)} \bigg\},\quad \text{ as  $n\to \infty$,}
\eeq
where $\Phi(x)$ is the cumulative distribution function for the standard normal random variable.
In particular,
\beq \label{decomp}
Q=A_n+B_n
\eeq
where for any small constant $\epsilon>0$, it holds that $\textup{cov}\{A_n/n,n^{-5/6}(B_n-2n^{1/2}\|u_1\|_1^2)\}=O(n^{-1/6+\epsilon})$ and
\beq \label{lim.dist}
\frac{A_n}{n}\to_d N\{0,2(1-2/\pi)^2\}, \quad\quad n^{-5/6}(B_n-2n^{1/2}\|u_1\|_1^2) \to_d \frac{2}{\pi} \mathcal{TW}_1,
\eeq
where $\mathcal{TW}_1$ is the Tracy-Widom distribution. 
\eet

\begin{remark}
	From the above theorem, the limit distribution of the normalized statistic $n^{-1}(Q-2n^{1/2}\|u_1\|_1^2)$ is normal $N\{0,2(1-2/\pi)^2\}$ as $n\to\infty$. In other words, for large $n$, the modularity $Q$ is roughly distributed around the center $2n^{1/2}\|u_1\|_1^2$ with the standard error $2^{1/2}(1-2/\pi)n$. In particular, for GOE, it can be shown that the functional $\|u_1\|_1^2$ of the first eigenvector $u_1$ satisfies $| \|u\|_1^2/n-2/\pi|=O_P(n^{-1/2}).$ As a result, one has $Q/n=4n^{1/2}\pi+O_P(1)$, which means $Q/n$ is concentrated around $4\pi n^{1/2}$, with a constant order fluctuation. 
\end{remark}

\begin{remark}
	From the second statement in Theorem \ref{GOE.thm}, we know that  $Q$ can be decomposed into two weakly dependent parts. The first part is asymptotically a centered normal, whereas the second part can be characterised by a shifted and scaled Tracy-Widom random variable. In particular, according to the characterization in (\ref{lim.dist}), the standardized statistic can be decomposed as
	\beq \label{standardized.Q}
	n^{-1}(Q-2n^{1/2}\|u_1\|_1^2)=\frac{A_n}{n}+\frac{B_n-2n^{1/2}\|u_1\|_1^2}{n},
	\eeq
	where $n^{-1}(B_n-2n^{1/2}\|u_1\|_1^2)=O(n^{-1/6}).$
	Hence the contribution to the asymptotic variance from $B_n$ diminishes as $n\to\infty$, at the rate of $n^{-1/6}$. In other words, the term $A_n$ is responsible for the asymptotic variance whereas the term $B_n$ only contributes to the asymptotic mean of $Q$.
\end{remark}

The proof of the above theorem relies on the key observation that
\beq
Q(W)=\lambda_1(W) \{\mbox{sgn}(u_1^\top)u_i\}^2+\sum_{i=2}^n \lambda_i(W_G) \{\mbox{sgn}(u_1^\top)u_i\}^2,
\eeq
where the two terms can be treated separately. In fact, by setting $B_n=\lambda_1(W) \{\mbox{sgn}(u_1^\top)u_i\}^2$ and $A_n=\sum_{i=2}^n \lambda_i(W_G) \{\mbox{sgn}(u_1^\top)u_i\}^2$, the statement (\ref{lim.dist}) in Theorem \ref{GOE.thm} can be proved by carefully analysing the joint distribution of the GOE eigenvalues, eigenvectors and their functionals. In particular, to show the asymptotic normality of $A_n/n$, we adopted several technical tools including the Haar measure on the orthogonal group $O(n)$, a Berry-Esseen bound for exchangeable pairs of random vectors, the semicircle law and the eigenvalue rigidity result for Wigner matrices. We leave the detailed proof of Theorem \ref{GOE.thm} to the Appendix.

\subsection{Second-Order Correction using Convolution}

Practically, as the variance contribution from $B_n$ diminishes at a very slow rate, it could be far from precise to use $N\{0,2(1-2/\pi)^2\}$ as an approximation of the empirical distribution of $n^{-1}(Q-2n^{1/2}\|u_1\|_1^2)$. Instead, by Theorem 1, we suggest taking the $n^{-1/6}$ order term into account and using the convolution of independent normal $ N\{0,2(1-2/\pi)^2\}$ and rescaled Tracy-Widom distribution $ {2}{n^{-1/6}\pi^{-1}} \mathcal{TW}_1$, as a finite sample approximation of the limiting distribution. Hereafter we denote the cumulative distribution function of the convolution as $F$.
The empirical performance of such convolutional approximation is assessed in Section \ref{simu.sec}.

Moreover, in Table \ref{table:t0}, we numerically evaluate the correlation between $A_n$ and $B_n$. The vanishing correlation in this case provides another justification of our use of convolution for the second-order approximation.
\begin{table}[h!]
	\centering
	\caption{Empirical correlation between $A_n$ and $B_n$. Each correlation is estimated over $10^5$ iterations.}
	\begin{tabular}{c|cccc}
		\hline
		n &   50  & 100 & 500 & 1000\\
		\hline
		$\mbox{cor}(A_n,B_n)$ & 0$\cdot$021  &   0$\cdot$016 &  0$\cdot$005  & 0$\cdot$003\\
		\hline
	\end{tabular}
	\label{table:t0}
\end{table}

\subsection{Universality Implied by Random Matrix Theory}

Although the limit distribution (\ref{lim.law}) of Theorem 1 was proven under the standard GOE setting, the analysis only relies on the joint distribution of the eigenvalues and the eigenvectors of GOE, as yielded by the proof in our Appendix. In this section, we discuss the potential universality of our results, or its generalizability to other matrix ensembles.

In connection to the recent achievements in Random Matrix Theory, it has been shown that the asymptotic behaviour of the eigenvalues and eigenvectors of many important classes of random matrices are the same as those of the GOE. For example, the well-known semicircle law has been obtained for the sample covariance matrices \citep{bai1988necessary}, the sample correlation matrices \citep{jiang2004limiting}, the Erd{\H{o}}s-R\'enyi graphs \citep{erdHos2013spectral}, the random regular graphs \citep{bauerschmidt2017local}, the generalized Wigner matrices \citep{tao2010random,erdHos2012rigidity} and the deformed Wigner matrices \citep{knowles2013isotropic}; universality results for eigenvectors have been obtained for the generalized Wigner matrices \citep{tao2011random,knowles2013eigenvector,bourgade2017eigenvector} and, more recently, the sample covariance matrices \citep{bloemendal2016principal,ding2019singular}. 

For the matrix ensembles whose spectral behaviour deviates significantly from those of GOE, we admit that the same limit distribution (\ref{lim.law}) would not hold in general. For example, when $p/n\to \gamma \in(0,\infty)$, it is well known that the limiting eigenvalue distribution for the sample covariance matrices is a non-symmetric Marcenko-Pastur law \citep{marchenko1967distribution}. In this case, (\ref{lim.law}) become questionable as some of the calculations, such as Equation (\ref{alpha}) in Appendix, will no longer hold. However, we do want to emphasize that the analytical framework developed in this paper is generic, and can be  applied to derive the asymptotic distributions under other settings, although the calculation of some relevant quantities (such as those paralleling Lemma A1, A2, and A4) might be technically challenging.

\subsection{Comparison with $\lambda_1(W)$}

In addition to the modularity statistic studied in this paper, some other statistics have been proposed for the purpose of community detection, especially the largest eigenvalue $\lambda_1(W)$ \citep{bickel2016hypothesis,lei2016goodness}. The distribution of the largest eigenvalue of the Wigner matrix is well understood. \citet{tracy1994level} first derived this distribution, and given the ostensibly prominent role that $\lambda_1$ plays in $Q$, some have used the close relationship between $\lambda_1$ and $Q$ in order to test for the presence of community structure in networks \citep{bickel2016hypothesis,lei2016goodness}. Here we investigate how close of a proxy $\lambda_1$ is to $Q$ to see if this approximation is justified. To evaluate this, we consider the correlation $\mbox{cor}(Q/n,n^{1/6}\lambda_1)$. The following theorem provides a negative answer by showing the  asymptotic uncorrelatedness between $Q/n$ and $n^{1/6}\lambda_1$.

\bet
Under the condition of Theorem 1, it holds that, for any $\epsilon>0$,
\beq
\textup{cov}(Q/n,n^{1/6}\lambda_1)=O(n^{-1/6+\epsilon}).
\eeq
\eet

\begin{figure}
	\centering
	\includegraphics[angle=0,width=6in]{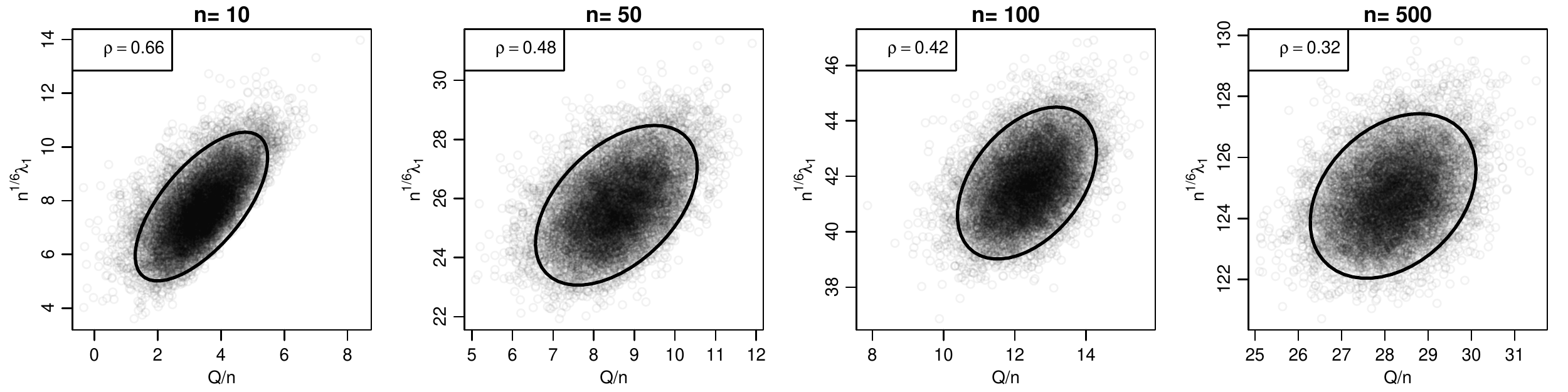}
	\caption{{Relationship between modularity and the largest eigenvalue of the modularity matrix.}  Standard Wigner matrices were generated 10,000 times for each $n$ to produce the scatter plots and corresponding correlation estimates. }
	\label{EigFig}
\end{figure}

In Figure \ref{EigFig}, we show the scatter plots of $n^{1/6}\lambda_1$ and $Q/n$ for various $n$, based on 10,000 simulations from a standard Wigner matrix as defined in Section~\ref{classicGOE:sec}. As a result, a clear decrease in the empirical correlation can be observed as $n$ increases, indicating the poor asymptotic approximation of $Q$ by $\lambda_1$.

\subsection{Statistical Power for Community Detection}

Our next result concerns the statistical power of the test based on the normalized modularity $n^{-1}(Q-2{n}^{1/2}\|u_1\|_1^2)$ and its limiting distribution obtained under the GOE null. Specifically, for a given signed edge weight matrix $W$, we calculate its first eigenvector $u_1$, and reject the null hypothesis whenever $n^{-1}(Q-2n^{1/2}\|u_1\|_1^2)>\Phi^{-1}(1-\alpha)$ for some desired level $\alpha\in(0,1)$. Naturally, one could replace $\Phi^{-1}(1-\alpha)$ by $F^{-1}(1-\alpha)$ for better finite sample performance. For the alternative model, we consider the following deformed GOE model where the signed edge weight matrix $W=\Theta+Z\in \R^{n\times n}$ is symmetric, with $\Theta$ incorporating the underlying community structure and $Z$ being a standard Wigner matrix. Examples of $\Theta$ include block-wise constant matrices or block-wise diagonal matrices extensively studied under the stochastic block models \citep{lei2015consistency,zhang2016minimax,hu2020using}, and some general low rank matrices commonly considered for studying the spectral clustering algorithms \citep{lu2016statistical,loffler2019optimality}.

\bet \label{power.thm}
Suppose $W=\Theta+Z\in \R^{n\times n}$, where $Z$ is a standard Wigner matrix and $\Theta$ is some fixed symmetric matrix. Then, as long as $\lambda_1(\Theta)\ge C_0\surd{n}$ and $\lambda_n(\Theta)>-C_1\surd{n}$ for some universal constants $C_0,C_1>0$, we have $\textup{pr}\{n^{-1}(Q-2n^{1/2}\|u_1\|_1^2)>\Phi^{-1}(1-\alpha)\}\to 1$ and $\textup{pr}\{n^{-1}(Q-2n^{1/2}\|u_1\|_1^2)>F^{-1}(1-\alpha)\}\to 1$.
\eet

\begin{remark}
	The alternative model considered in Theorem \ref{power.thm} covers a wide range of scenarios where community structure is present in the network edge weights. In particular, the theorem only requires the matrix $\Theta$ to have sufficiently large global signal to be detectable from the noisy observations, and there is no need to specify the community structure incorporated in $\Theta$. 
\end{remark}

\section{Simulations} \label{simu.sec}

\subsection{Empirical Quantile Assessment}

In this section, we conduct simulation studies to empirically evaluate our derived limiting distribution in the previous section. We generate the GOE matrices whose dimension $n$ varies from 50 to 5,000, and calculate their modularities defined by (\ref{eqmod}). 
We compare the empirical distributions of the standardized statistic in (\ref{standardized.Q}) based on simulated modularities against their theoretical quantiles $q_\alpha$ corresponding to probabilities $\alpha$ varying from 0$\cdot$01 to 0$\cdot$99. Specifically, we evaluate the tail probabilities of the standardized statistic at $q_\alpha$ for different $n$ and cutoffs $\alpha$. For the theoretical distribution, we consider two distributions, namely, the normal distribution $N\{0,2(1-2/\pi)^2\}$ obtained in Theorem \ref{GOE.thm}, and its second-order correction $F$. In particular, in the latter case, the theoretical quantiles are obtained numerically by generating 100,000 samples. Table \ref{table:t1} and Table \ref{table:t2} show the empirical quantiles based on 100,000 rounds of simulations. Comparing these two tables, it is clear that the convolution $F$ provides a better approximation of the empirical distribution.

\begin{table}[h!]
	\centering
	\caption{Empirical probabilities at $N\{0,2(1-2/\pi)^2\}$ quantiles: the GOE case}
	\begin{tabular}{c|cccccc}
		\hline
		$n$ & 50 &  100  & 500 & 1000 & 2000  & 5000\\
		\hline
		$\alpha=0\cdot$01 & $0\cdot$0061 &  $0\cdot$0054 &  $0\cdot$0059 & $0\cdot$0056 & $0\cdot$0056 &  $0\cdot$0057  \\
		
		$\alpha=0\cdot$05 & $0\cdot$0210  &  $0\cdot$0226 & $0\cdot$0251 & $0\cdot$0260 & $0\cdot$0268 &  $0\cdot$0283 \\
		
		$\alpha=0\cdot$25 & $0\cdot$0914 &  $0\cdot$1042 & $0\cdot$1295 & $0\cdot$1364 & $0\cdot$1470 & $0\cdot$1602\\
		
		$\alpha=0\cdot$50 & $0\cdot$2033 &$0\cdot$2313 & $0\cdot$2927 & $0\cdot$3104 & $0\cdot$3322 &  $0\cdot$3564\\
		
		$\alpha=0\cdot$75 & $0\cdot$3745 &  $0\cdot$4221 &  $0\cdot$5162 & $0\cdot$5469 & $0\cdot$5743 & $0\cdot$6054\\
		
		$\alpha=0\cdot$95 & $0\cdot$6668 & $0\cdot$7231 &  $0\cdot$8134 & $0\cdot$8397 & $0\cdot$8588 &  $0\cdot$8808\\
		
		$\alpha=0\cdot$99 & $0\cdot$8337 &  $0\cdot$8758 &  $0\cdot$9326 & $0\cdot$9455 & $0\cdot$9551 & $0\cdot$9652\\
		\hline
	\end{tabular}
	\label{table:t1}
\end{table}

\begin{table}[h!]
	\centering
	\caption{Empirical probabilities at $F$ quantiles: the GOE case}
	\begin{tabular}{c|cccccc}
		\hline
		$n$ & 50 & 100 &  500 & 1000 & 2000 & 5000\\
		\hline
		$\alpha=0\cdot$01 & $0\cdot$0191 & $0\cdot$0154 & $0\cdot$0125 & $0\cdot$0128 & $0\cdot$0103 & $0\cdot$0114  \\
		
		$\alpha=0\cdot$05 & $0\cdot$0813 & $0\cdot$0709 &  $0\cdot$0580 & $0\cdot$0570 & $0\cdot$0534 &  $0\cdot$0526\\
		
		$\alpha=0\cdot$25 & $0\cdot$1835 & $0\cdot$2016 &  $0\cdot$2293 & $0\cdot$2350 & $0\cdot$2353 &  $0\cdot$2427\\
		
		$\alpha=0\cdot$50 & $0\cdot$4078 & $0\cdot$4314 & $0\cdot$4689 & $0\cdot$4804 & $0\cdot$4886 & $0\cdot$4889\\
		
		$\alpha=0\cdot$75 & $0\cdot$6720 & $0\cdot$6922 &  $0\cdot$7257 & $0\cdot$7320 & $0\cdot$7464 &  $0\cdot$7405\\
		
		$\alpha=0\cdot$95 & $0\cdot$9179 & $0\cdot$9286 &  $0\cdot$9432 & $0\cdot$9448 & $0\cdot$9469 &  $0\cdot$9482\\
		
		$\alpha=0\cdot$99 & $0\cdot$9808 & $0\cdot$9843 & $0\cdot$9891 & $0\cdot$9882 & $0\cdot$9889 & $0\cdot$9888\\
		\hline
	\end{tabular}
	\label{table:t2}
\end{table}

\subsection{Universality of the Limit Distribution}

In this section, we empirically evaluate the validity of our theoretical limit distribution as well as its second-order approximation under some non-GOE matrix ensembles. In particular, as of both theoretical and practical interest, we consider (i) symmetric random matrices with heavy-tailed, nonsymmetric distributions such as the exponential distribution, Exp(1); (ii) the adjacency matrix of sparse Erd{\H{o}}s-R\'enyi random graph \citep{erdHos2012spectral,erdHos2013spectral} with $p=n^{-1/4}$; and (iii) the sample correlation matrix $R_n$ of $N$ independent observations from $N(0,I_n)$ with $N=n^{5/2}$. In case (i) and (ii), the entries of the random matrices are normalized to match the first two moments of GOE. In case (iii), the modularity is calculated from the normalized matrix ${N^{1/2}}(R_n-I_n)$. In Table \ref{table:t4}-\ref{table:t7}, we show the empirical tail probabilities evaluated at different quantiles of the convolution $F$. The results concerning tail probabilities evaluated at the quantiles of $N\{0,2(1-2/\pi)^2\}$ are put in our Supplementary Material. Our numerical results suggest the universality of our limit distribution as well as its second-order approximation over a wide range of non-GOE random matrices, which implies its strong potential for practical applications beyond the GOE setting.

\begin{table}[h!]
	\centering
	
	\caption{Empirical probabilities at $F$ quantiles: heavy-tailed nonsymmetric distribution Exp(1)}
	\begin{tabular}{c|cccccc}
		\hline
		$n$ & 50 & 100 & 500 & 1000 & 2000  & 5000\\
		\hline
		$\alpha=0\cdot$01 & $0\cdot$0884 & $0\cdot$0796 &  $0\cdot$0253 & $0\cdot$0197 & $0\cdot$0155 & $0\cdot$0130  \\
		
		$\alpha=0\cdot$05 & $0\cdot$1900 & $0\cdot$1839 &  $0\cdot$0957 & $0\cdot$0795 & $0\cdot$0689 &  $0\cdot$0616 \\
		
		$\alpha=0\cdot$25 & $0\cdot$4387 & $0\cdot$4572 &  $0\cdot$3513 & $0\cdot$3204 & $0\cdot$3002 &  $0\cdot$2800\\
		
		$\alpha=0\cdot$50 & $0\cdot$6479 & $0\cdot$6790 &  $0\cdot$6086 & $0\cdot$5825 & $0\cdot$5570 &  $0\cdot$5338\\
		
		$\alpha=0\cdot$75 & $0\cdot$8271 & $0\cdot$8540 & $0\cdot$8236 & $0\cdot$8078 & $0\cdot$7920 & $0\cdot$7773\\
		
		$\alpha=0\cdot$95 & $0\cdot$9615 & $0\cdot$9732 &  $0\cdot$9689& $0\cdot$9666 & $0\cdot$9619 & $0\cdot$9595 \\
		
		$\alpha=0\cdot$99 & $0\cdot$9912 & $0\cdot$9950 & $0\cdot$9948 & $0\cdot$9937 & $0\cdot$9928 &  $0\cdot$9920 \\
		\hline
	\end{tabular}
	\label{table:t4}
\end{table}

\begin{table}[h!]
	\centering
	\caption{Empirical probabilities at $F$ quantiles: sparse Erd{\H{o}}s-R\'enyi random graph ($p=n^{-1/4}$)}
	\begin{tabular}{c|cccccc}
		\hline
		$n$ & 50 & 100 & 500 & 1000 & 2000  & 5000\\
		\hline
		$\alpha=0\cdot$01 & $0\cdot$0006 & $0\cdot$0015 &  $0\cdot$0067 & $0\cdot$0094 & $0\cdot$0110 & $0\cdot$0128  \\
		
		$\alpha=0\cdot$05 & $0\cdot$0059 & $0\cdot$0121 &  $0\cdot$0374 & $0\cdot$0465 & $0\cdot$0557 &  $0\cdot$0606 \\
		
		$\alpha=0\cdot$25 & $0\cdot$0681 & $0\cdot$1085 &  $0\cdot$2093 & $0\cdot$2424 & $0\cdot$2646 &  $0\cdot$2784\\
		
		$\alpha=0\cdot$50 & $0\cdot$2261 & $0\cdot$3035 &  $0\cdot$4464 & $0\cdot$4921 & $0\cdot$5181 &  $0\cdot$5344 \\
		
		$\alpha=0\cdot$75 & $0\cdot$4958 & $0\cdot$5776 & $0\cdot$7069 & $0\cdot$7403 & $0\cdot$7648 & $0\cdot$7759 \\
		
		$\alpha=0\cdot$95 & $0\cdot$8524 & $0\cdot$8899 &  $0\cdot$9354& $0\cdot$9458 & $0\cdot$9541 & $0\cdot$9562 \\
		
		$\alpha=0\cdot$99 & $0\cdot$9633 & $0\cdot$9733 & $0\cdot$9870 & $0\cdot$9894 & $0\cdot$9913 &  $0\cdot$9913 \\
		\hline
	\end{tabular}
	\label{table:t6}
\end{table}

\begin{table}[h!]
	\centering
	\caption{Empirical probabilities at $F$ quantiles: sample correlation matrix ($N=n^{5/2}$)}
	\begin{tabular}{c|cccccc}
		\hline
		$n$ & 20  &  50 & 75 & 100 & 150 & 200 \\
		\hline
		$\alpha=0\cdot$01 & $0\cdot$0033 & $0\cdot$0068 &  $0\cdot$0059 & $0\cdot$0073 & $0\cdot$0076 & $0\cdot$0083  \\
		
		$\alpha=0\cdot$05 & $0\cdot$0176 & $0\cdot$0310 &  $0\cdot$0343 & $0\cdot$0383 & $0\cdot$0480 &  $0\cdot$0465 \\
		
		$\alpha=0\cdot$25 & $0\cdot$1280 & $0\cdot$1791 &  $0\cdot$2075 & $0\cdot$2186 & $0\cdot$2356 &  $0\cdot$2373  \\
		
		$\alpha=0\cdot$50 & $0\cdot$3289 & $0\cdot$4156 &  $0\cdot$4561 & $0\cdot$4648 & $0\cdot$4880 &  $0\cdot$4905 \\
		
		$\alpha=0\cdot$75 & $0\cdot$6164 & $0\cdot$6914 & $0\cdot$7219 & $0\cdot$7308 & $0\cdot$7479 & $0\cdot$7550 \\
		
		$\alpha=0\cdot$95 & $0\cdot$9197 & $0\cdot$9364 &  $0\cdot$9450 & $0\cdot$9491 & $0\cdot$9532 & $0\cdot$9516 \\
		
		$\alpha=0\cdot$99 & $0\cdot$9829 & $0\cdot$9872 & $0\cdot$9912 & $0\cdot$9896 & $0\cdot$9919 &  $0\cdot$9908 \\
		\hline
	\end{tabular}
	\label{table:t7}
\end{table}

\subsection{Empirical Power Assessment}

Under certain alternative model for weighted signed networks that suggest community structure, we numerically assess and compare powers of the modularity based tests and some other methods for community detection. Specifically, we consider the following deformed/spiked GOE model where the signed edge weight matrix $W=\beta uu^\top+D+Z\in \R^{n\times n}$ is symmetric, with $\beta\in \R$, $u\in S^{n-1}$, $D$ a diagonal matrix, and $Z$ a standard Wigner matrix. In particular, for each $n\in \{100,200,300,400,500,600\}$, we set $\beta=\surd{n}$, and set $u$ such that its first $n/2$ coordinates are $n^{-1/2}$ and the rest of the coordinates are $-n^{-1/2}$. This implies two clusters of nodes of equal size, where the within-group  and cross-group edge weights are two distinct values. The diagonal entries of $D$ are randomly generated from $[-\surd{n},\surd{n}]$, to increase heterogeneity. Table \ref{table:t8} shows the empirical powers of (i) Modularity Test I: the test based on the normalized modularity and its Gaussian limiting distribution in Theorem \ref{GOE.thm}, (ii) Modularity Test II: the test based on the normalized modularity and the convolutional approximation $F$, whose quantiles are obtained numerically as in previous sections, (iii) Largest Eigenvalue Test: the test based on $\lambda_1(W)$ and its Tracy-Widom limiting distribution \citep{johnstone2012fast}, and (iv) Entrywise Maximum Test: the test based on the entrywise maxima $\max_{1\le i\ne j\le n}|\{\text{cov}(W)\}_{ij}|$ and its Gumbel limiting distribution \citep{jiang2004asymptotic,hu2020using}. The details of the Largest Eigenvalue Test and the Entrywise Maximum Test and their asymptotic validity under the null model are demonstrated in our Supplementary Material. The empirical powers of these methods at level $\alpha=0.05$ are calculated from 100,000 rounds of simulations. From Table \ref{table:t8}, we find that the Modularity Tests I and II are more powerful than the Largest Eigenvalue Test and the Entrywise Maximum Test when $n$ is large ($n\ge 400$), while the Entrywise Maximum Test is more powerful for smaller $n$.

\begin{table}[h!]
	\centering
	\caption{Empirical powers of four different methods at level $\alpha=0.05$}
	\begin{tabular}{c|cccccc}
		\hline
		$n$ & 50  &  100 & 200 & 400 & 600 & 800 \\
		\hline
		Modularity Test I  & $0\cdot$3791 & $0\cdot$4779 &  $0\cdot$5819 & $0\cdot$6653 & $0\cdot$7106 & $0\cdot$7358  \\
		
		Modularity Test II & $0\cdot$4432 & $0\cdot$5569 &  $0\cdot$6493 & $0\cdot$7164 & $0\cdot$7525 &  $0\cdot$7793 \\
		
		Largest Eigenvalue Test & $0\cdot$5471 & $0\cdot$5804 &  $0\cdot$6159 & $0\cdot$6510 & $0\cdot$6788 &  $0\cdot$6952 \\
		
		Entrywise Maximum Test & $0\cdot$7686 & $0\cdot$7108 &  $0\cdot$6655 & $0\cdot$6206 & $0\cdot$5977 &  $0\cdot$5969 \\
		\hline
	\end{tabular}
	\label{table:t8}
\end{table}

\section{Real Data Analysis}

\subsection{Analysis of US Congressional Voting Networks}

Annual voting records for individuals in the U.S. house of representatives provides a commonly used example of a highly modular network, where nodes stand for representatives and edge weights correspond with the correlation between the voting records of pairs of representatives. Recently, evidence of increased partisan polarization has been observed based on increased modularity in more recent annual voting networks \citep{neal2018sign}. We let $W$ be a centered and scaled correlation matrix with zeroes in the diagonal based on the 1984 congressional voting records. We removed congressmen with more than 50\% votes unrecorded for the year for a total of $n=431$ congressmen in our network. Letting $\mbox{sgn}(u_1)$ determine community membership, representatives were strongly divided based on party affiliation, with 96$\cdot$9\% Democrat and 3$\cdot$1\% Republican membership in one community, and 77$\cdot$6\% Republican and 22$\cdot$4\% Democrat membership in the other community. Modularity was very large, with $Q/n-2\|u_1\|_1^2/n^{1/2} = 314$$\cdot$3, which based on Theorem \ref{GOE.thm} provides overwhelming statistical evidence of community structure.

Given the nature of partisan politics, it is unsurprising that there was strong evidence to reject a null hypothesis of no community structure in congress. Thus we also explore the less obvious question of whether there are additional communities beyond the Republican and Democrat divide. By restricting the data to the 205 congressmen in the Republican-dominated community, namely, 77$\cdot$6\% republican, we recentered and rescaled weights over this subset and applied Theorem \ref{GOE.thm} again. There again was overwhelming statistical evidence of additional community structure as $Q/n-2\|u_1\|_1^2/n^{1/2} = 98$$\cdot$5. The largest eigenvalue and the entrywise maximum-based tests led to consistent conclusions with p-values less than $1$$\cdot$$0 \times 10^{-4}$ in both cases. Therefore, this Republican dominated subset divides into two additional communities: one community with a majority, 58$\cdot$7\%, of Democrats, and another with a large majority, 88$\cdot$1\%, of Republicans. This is evidence of a substantial subset of moderate Democrats that, despite being initially clustered with Republicans based on their voting record, also demonstrated sufficient differences from the Republicans to warrant belonging to a separate and distinct community.

\subsection{Network Structure of the Human Cranium }

\begin{figure}
	\centering
	\includegraphics[angle=0,width=11cm]{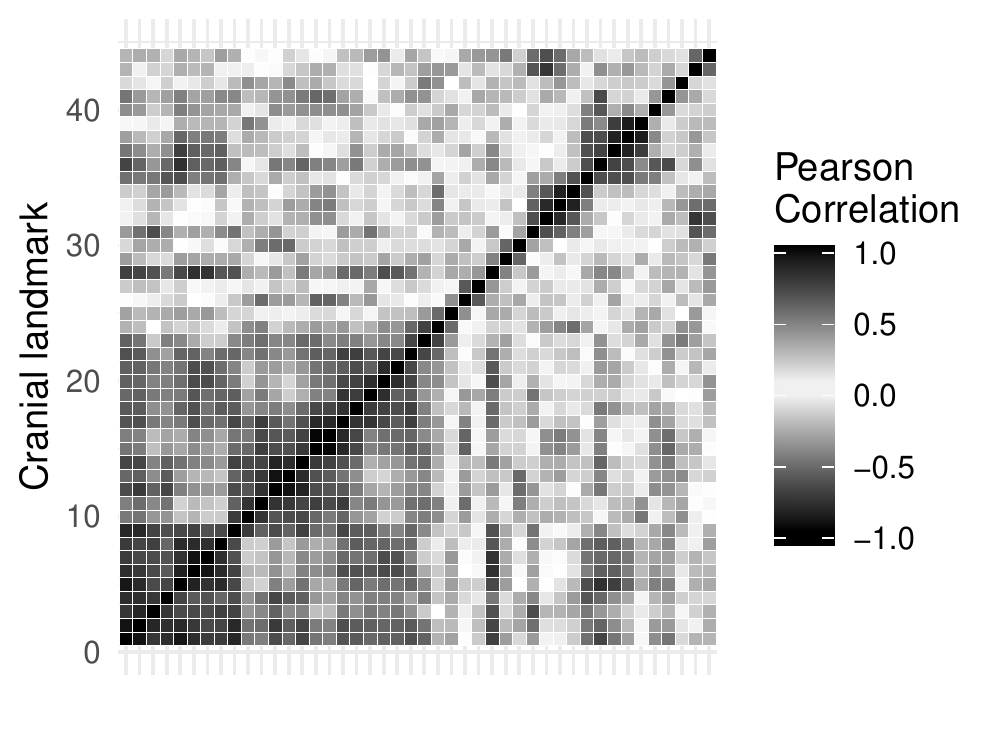}
	\caption{{Correlation network of landmark measurements of the human cranium.}  Cranial landmarks are discrete anatomical points that are homologous across humans. A sample of 1,367 male crania were to used to calculate the Pearson correlations between each pair of 44 cranial landmark measurements.   }
	\label{CraniumFig}
\end{figure}

Morphological networks of the human cranium define nodes to be anatomically defined measurements between landmark points on the cranium of a particular individual. Edge weights are defined by Pearson correlations between cranial measurements for each pair of landmarks. In Fig.~\ref{CraniumFig}, the corresponding correlation network demonstrates blocks of cranial landmarks with nested correlation structure, such as what can be observed for landmarks 1 through 24. Due to different cranial landmarks developing simultaneously on the cranium for each individual, and therefore subject to the same environmental factors throughout development,  this nested structure is an expected feature of this morphological network. Network nestedness occurs when interactions of less connected nodes form proper subsets of the interactions of more connected nodes. Modularity is a type of nestedness where there is no distinct heirarchical structure separating nodes with low degree from nodes with high degree within a community, and so modularity can be interpreted as an intermediate form of nestedness.

\citet{cantor2017nestedness} constructed morphological networks of the human crania using 1,367 males to calculate correlations between each pair of 44 different landmark measurements. They found significant statistical evidence of nestedness in the resulting correlation network. 
We apply Theorem \ref{GOE.thm} to this network after proper normalization and find that there is overwhelming statistical evidence of community structure ($Q/n-2\|u_1\|_1^2/n^{1/2} = 25\cdot$7).
Similarly, tests for the same null hypothesis based on the asymptotic distributions of the first eigenvalue and the entrywise maximum both lead to the same conclusion with both p-values less than $1$$\cdot$$0 \times 10^{-4}$. This implies that human crania tend to contain clusters of landmarks, likely spatially close to one another, that grow together in parallel throughout development.

\section{Discussion}

Our numerical results show that, although having a significant improvement upon the original limit distribution, the second-order approximation seems still insufficient for applications with small sample sizes. Hence, it would be interesting to find some more accurate higher-order approximation for the limit distribution in Theorem \ref{GOE.thm}.

In \cite{reichardt2006networks} and \cite{fortunato2007resolution}, it was shown that the modularity defined as in Equation (\ref{eqmod}) has its own limits, such as it is unable to find community structure in networks with many small communities. To address the issue, \cite{arenas2008analysis} proposed a generalized modularity which includes a resolution parameter. Consequently, it would also be of interest to extend our analysis to the generalized modularity \citep{newman2016equivalence}.

In Section 2.1, due to complicated dependence structure between the error term $n^{1/2}(\|u\|_1^2/n-2/\pi)$ and the first order fluctuation $A_n/n$, our current analytical framework can only lead us to the limiting distribution of the normalized $n^{-1}(Q-2n^{1/2}\|u_1\|_1^2)$. We admit this is mainly due to the limitation of our technical tools, and, in light of Remark 1, it is of interest whether a direct limiting distribution for $n^{-1}(Q-n^{3/2}4/\pi)$ can be obtained. Some numerical comparisons of $n^{-1}Q$, $n^{-1}(Q-2n^{1/2}\|u_1\|_1^2)$ and $n^{-1}(Q-n^{3/2}4/\pi)$ are presented in our Supplementary Material, which suggest that a test based on $n^{-1}(Q-n^{3/2}4/\pi)$ could be more powerful against certain alternatives. We leave the more rigorous theoretical investigations for future research.

\section*{Acknowledgement}
Research reported in this publication was supported by the National Institute Of Mental Health of the National Institutes of Health under Award Number R01MH116884 (IB). 

\appendix
\section*{Appendix}

Throughout, for sequences $\{a_n\}$ and $\{b_n\}$, we write $a_n = o(b_n)$ (or $a_n = o_P(b_n)$) if $\lim_{n} a_n/b_n =0$ (in probability), and write $a_n = O(b_n)$ (or $a_n = O_P(b_n)$), $a_n\lesssim b_n$ or $b_n \gtrsim a_n$ if there exists a constant $C$ such that $a_n \le Cb_n$ for all $n$ (in probability). We write $a_n\asymp b_n$ if $a_n \lesssim b_n$ and $a_n\gtrsim b_n$. For a set $A$, we denote $|A|$ as its cardinality. Lastly, $C, C_0, C_1,...$ are constants that may vary from place to place.

In the following, we prove Theorem 1 in the main paper. The proofs of other theorem and some technical lemmas are collected in our online Supplementary Material.

\emph{Proof of Theorem 1.}
We first recall some important results concerning the eigenvectors and the eigenvalues of GOE. Specifically, the eigenvectors $u_1(W),...,u_n(W)$ are uniformly distributed on the half-sphere $S_+^{n-1} = \{x=(x_1,...,x_n)\in S^{n-1}: x_1>0)$,
and the joint distribution of $(u_1(W),...,u_n(W))$ is the Haar measure on the orthogonal group $O(n)$, with each column multiplied by $-1$ or $1$ so that the columns all belong to $S_+^{n-1}$ \citep{o2016eigenvectors}. An immediate consequence is the following proposition characterizing the joint distribution of the eigenvectors of $W$.

\bep  \label{vector.dist}
Let $v$ be a random vector uniformly distributed on  $S^{n-1}$. Then $v$ has the same distribution as $\big( \xi_1(\sum_{j=1}^n \xi_j^2)^{-1/2}, ...,\xi_n(\sum_{j=1}^n \xi_j^2)^{-1/2} \big)$
where $\xi_1,...,\xi_n$ are independently and identically drawn from $N(0,1)$.
\eep
Another well-known fact related to the modularity under GOE is the limiting distribution of its largest eigenvalue $\lambda_1(W)$, derived in the seminal works of Tracy and Widom \citep{tracy1994level,tracy1996orthogonal}.

\bet[The Tracy-Widom Law] \label{GOE.TW}
Let $\lambda_1(W)$ denote the largest eigenvalue of $W$ where $W$ is a sample from GOE with dimension $n\times n$, then $ \textup{pr}\{ n^{1/6}(\lambda_1(W)-2n^{1/2}) \le s\}\to F(s)$, where $F(s)$ denotes the Tracy-Widom distribution.
\eet

\noindent{\emph{Proof of Theorem \ref{GOE.thm}}.} By definition and eigen-decomposition of $W$, we have
\begin{align} \label{Qn.eq0}
Q/n&=\textup{sgn}(u_{1}^\top)W \textup{ sgn}(u_{1})/n =\lambda_1(W)\{\textup{sgn}(u_{1}^\top)u_1\}^2/n+ \sum_{i=2}^n\lambda_i(W)\{\textup{sgn}(u_{1}^\top)u_i\}^2/n \nonumber \\
&=\lambda_1(W)\|u_1\|_1^2/n+ \sum_{i=2}^n\lambda_i(W)\{\textup{sgn}(u_{1}^\top)u_i\}^2/n\equiv B_n/n+A_n/n.
\end{align}
The proof is separated into three parts. Firstly, we show that $B_n/n$ in (\ref{Qn.eq0}), after proper centring and scaling, converges weakly to a Tracy-Widom distribution. Secondly, we show that $A_n/n$ is asymptotically normal. Finally, we deal with the covariance between the two terms.

\bigskip
\noindent{\bf Part I.}
By Proposition \ref{vector.dist}, we know that $u_1(W),u_2(W),...,u_n(W)$ are independent and have the same distribution as 
\beq \label{g.rep}
\bigg( \frac{\xi_1}{\surd{\sum_{j=1}^n \xi_j^2}}, ...,\frac{\xi_n}{\surd{\sum_{j=1}^n \xi_j^2}} \bigg),
\eeq
where $\xi_1,...,\xi_n$ are i.i.d. standard normal random variables. Therefore, 
\[
\|u_1\|_1^2/n = \frac{(\sum_{j=1}^n|\xi_j|)^2}{n{\sum_{j=1}^n \xi_j^2}}= \frac{n}{{\sum_{j=1}^n \xi_j^2}}\cdot \bigg( \frac{\sum_{j=1}^n|\xi_j|}{{{n}}}\bigg)^2.
\]
On the one hand, note that $\xi_j^2$ are independent $\chi^2$ random variables, which satisfies sub-exponential tail bound. By standard concentration inequality for sub-exponential random variables such as Proposition 5.16 in \cite{vershynin2010introduction}, we have, for any $\epsilon>0$
\[
\textup{pr}\bigg\{ \bigg|\frac{1}{n}\sum_{i=1}^n \xi_i^2 - E( \xi_i^2)\bigg| > \bigg(\frac{\log \frac{1}{\epsilon}}{n}\bigg)^{1/2} \bigg\} < \epsilon^{c},
\]
for some constant $c>0$. On the other hand, standard concentration inequality for sub-gaussian random variables yields, for any $\epsilon>0$,
\[
\textup{pr}\bigg\{ \bigg|\frac{1}{n}\sum_{i=1}^n |\xi_i| - E (|\xi_i|)\bigg| > \bigg(\frac{\log \frac{1}{\epsilon}}{n}\bigg)^{1/2} \bigg\} < \epsilon^{c},
\]
for some constant $c>0$. Thus with probability at least $1-O(\epsilon^{c})$ for some $c>0$,
\begin{align} \label{u1/n}
\bigg| \frac{n}{{\sum_{j=1}^n \xi_j^2}}\cdot \bigg( \frac{\sum_{j=1}^n|\xi_j|}{{{n}}}\bigg)^2-{\frac{2}{\pi}}\bigg| &\le \bigg|\frac{n}{{\sum_{j=1}^n \xi_j^2}}-1\bigg|\cdot \frac{2}{\pi}+  \frac{n}{{\sum_{j=1}^n \xi_j^2}} \bigg|\bigg( \frac{\sum_{j=1}^n|\xi_j|}{{{n}}}\bigg)^2- (\E |\xi_i|)^2\bigg| \nonumber\\
&\le \frac{2}{\pi} \bigg(\frac{\log \frac{1}{\epsilon}}{n}\bigg)^{1/2} + 2\bigg(\frac{2}{\pi}\bigg)^{1/2}\cdot \bigg\{{\frac{\log \frac{1}{\epsilon}}{n}} +\bigg(\frac{\log \frac{1}{\epsilon}}{n}\bigg)^{1/2} \bigg\} \nonumber\\
&\le C(n^{-1}\log {\epsilon}^{-1})^{1/2}.
\end{align}
By Theorem 4, we have
\begin{align} \label{term1}
n^{1/6}\{\lambda_1(W)\|u_1\|_1^2/n-2\|u_1\|_1^2/n^{1/2}\}\to_d \frac{2}{\pi}\mathcal{TW}_1.
\end{align}
In other words, $\lambda_1(W)\|u_1\|_1^2/n-2\|u_1\|_1^2/n^{1/2} = O_P(n^{-1/6}).$

\bigskip
\noindent{\bf Part II.}
We denote the second term in (\ref{Qn.eq0}) as
\[
A_n/n=n^{-1}\sum_{i=2}^n\lambda_i(W)\{\textup{sgn}(u_{1}^\top)u_i\}^2.
\] 
Denote $\gamma_j$ for $j=1,...,n,$ as the classical location of the $j$-th eigenvalue (scaled by $n^{1/2}$) under the semicircle law ordered in increasing order. In other words,
\[
n\int_{-\infty}^{\gamma_j} \rho_{sc}(x)dx=j,\quad j=1,...,n,
\]
where $\rho_{sc}(x)=(2\pi)^{-1}\surd{(4-x^2)_+}$ is the semicircle law. Define
\beq \label{Omega0}
\Omega_0 = n^{-1/2}\sum_{i=2}^n\gamma_i\{\textup{sgn}(u_{1}^\top)u_i\}^2.
\eeq
In what follows, we show that $\Omega_0$ is asymptotically normal with variance $2(1-2/\pi)^2$, and then conclude by verifying $|\Omega_0-A_n/n|\to 0$ in probability.

\bigskip
\noindent{\bf Asymptotic normality of $\Omega_0.$} The proof of asymptotic normality depends on the following key observations about a single $\textup{sgn}(u_{1}^\top)u_i$.

\bel \label{norm.Delta}
Suppose $(u_1,...,u_n)$ has a Haar measure on orthogonal group $O(n)$. Then for any $i=2,...,n$, it holds that $\textup{sgn}(u_{1}^\top)u_i \to_d N(0,1-2/\pi).$ In particular, we have $\textup{sgn}(u_{1}^\top)u_i=\mathcal{N}_i+O_p(\log n/n^{1/2})$, where $\mathcal{N}_i$ are drawn independently from $N(0,\sigma^2)$ and $\sigma^2=1-2/\pi+o(1)$.
\eel

Our next result concerns the relation between two elements $\text{sgn}(u_1^\top)u_i$ and $\text{sgn}(u_1^\top)u_j$ where $i,j\in\{2,...,n\}, i\ne j$. In particular, we show that $( \text{sgn}(u_1^\top)u_i,\text{sgn}(u_1^\top)u_j)$ is an isotropic vector.

\bel \label{ind.Delta}
Suppose $(u_1,...,u_n)$ has a Haar measure on orthogonal group $O(n)$. Then for any $i,j\in \{2,...,n\}$ with $i\ne j$, it holds that $E \{\textup{sgn}(u_1^\top)u_i \textup{sgn}(u_1^\top)u_j\}=0$ and $E[\{\textup{sgn}(u_1^\top)u_i\}^2 \{\textup{sgn}(u_1^\top)u_j\}^2]=(1-2/\pi)^2+o(1).$
\eel

Now without loss of generality we assume $n$ is even, namely, $n=2m$ for some integer $m>0$. Define
\[
\gamma_i\{\textup{sgn}(u_{1}^\top)u_i\}^2=\zeta_i^2,\quad \text{ for $i=1,...,m$}
\]
and
\[
\gamma_{n-i+1}\{\textup{sgn}(u_{1}^\top)u_{n-i+1}\}^2=-\eta_i^2, \quad \text{ for $i=1,...,m$.}
\]
Hence
\[
\Omega_0 = n^{-1/2}\sum_{i=2}^n\gamma_i\{\textup{sgn}(u_{1}^\top)u_i\}^2=m^{-1/2}\sum_{i=2}^m (\zeta_i^2-\eta_i^2)/2^{1/2}+n^{-1/2}\gamma_n\{\textup{sgn}(u_{1}^\top)u_n\}^2.
\]
Set $\alpha_i=(\zeta_i^2-\eta_i^2)/2^{1/2}$ for $i=2,...,m$. It is easy to check
\beq \label{alpha}
E(\alpha_i)=0,\quad E(\alpha_i\alpha_j)=0,
\eeq
suing Lemma A2, the exchangeable property of the Haar measure on $O(n)$ and the symmetry $\gamma_i=-\gamma_{n-i+1}$. The asymptotic normality of $\Omega_0$ can be obtained from the following central limit theorem for the symmetric isotropic random vectors and the fact that, by Lemma \ref{norm.Delta}, $n^{-1/2}\gamma_n\{\textup{sgn}(u_{1}^\top)u_n\}^2\to 0$.

\bel \label{clt.lem}
Suppose $X=(X_1,...,X_n)\in \R^n$ has a distribution that is invariant under reflections in the coordinate hyperplanes and
\[
E (X_i)=0,\quad E (X_i^2)=\sigma_i^2<\infty,\quad E (X_iX_j)=0
\]
for $i,j\in \{1,...,n\}$ and $i\ne j$. Let $\theta=(\theta_1,...,\theta_n)\in S^{n-1}$ be a fixed vector and $\sigma_\theta^2=\sum_{i=1}^n\theta_i^2\sigma_i^2$. Then
\begin{align*}
\sup_{t\in \R}| \textup{pr}(\sum_{i=1}^n\theta_iX_i\le \sigma_\theta t)-\Phi(t)|&\le 2\{\sigma_\theta^{-4}\sum_{i,j}\theta_i^2\theta_j^2E (X_i^2X_j^2) -1\}^{1/2}\\ &\quad+({8}/{\pi})^{1/4}[{\sigma_\theta^{-3}}\{\max_{i}E(|X_i|^3)\}\sum_{i=1}^n|\theta_i|^3]^{1/2}.
\end{align*}
\eel

\bel \label{alpha.mom}
For all $i=2,...,m,$ it holds that $E (\alpha_i^2) = 2\gamma_i^2(1-2/\pi)^2+o(1)$. For any fixed $i,j\in\{2,...,m\}$ and $i\ne j$, it holds that $\textup{cov}(\alpha_i^2,\alpha_j^2) =o(1).$
\eel

Let $\Omega_0= \theta^\top \alpha+O(n^{-1/2})$, where $\theta=(1/\surd{m},...,1/\surd{m})^\top$ and $\alpha=(\alpha_2,...,\alpha_m)^\top$. Denote $\sigma_i^2=\E \alpha_i^2$.
It then follows that $\sigma^2_\theta=\frac{1}{m}\sum_{i=2}^m\sigma_i^2\equiv \sigma_{sum}^2/m$.
Combining Lemma \ref{clt.lem} and \ref{alpha.mom}, we have
\begin{align*}
&\quad\sup_{t\in \R}\bigg|\text{pr}\bigg(\frac{ \theta^\top \alpha}{m^{-1/2}\sigma_m}\le t\bigg) -\Phi(t)\bigg|\\
&\le 2m\sigma_{sum}^{-2}\bigg(\frac{1}{m^2}\sum_{2\le i,j\le m}\E \alpha_i^2\alpha_j^2-\frac{1}{m^2}\sum_{2\le i,j\le m}\sigma_i^2 \sigma_j^2\bigg)^{1/2}+Cm^{-1/4}\\
&=\frac{2m}{\sigma^2_{sum}}\bigg\{\frac{1}{m^2}\sum_{i=2}^m\text{Var}(\alpha_i^2)+\frac{1}{m^2}\sum_{2\le i\ne j\le m}\text{cov}(\alpha_i^2,\alpha_j^2)\bigg\}^{1/2}+Cm^{-1/4}
\end{align*}
for some constant $C>0$.
Now note that
\beq \label{conv.sigma.m}
\lim_{m\to\infty}\frac{\sigma_{sum}^2}{m}=\lim_{m\to\infty}\frac{2(1-2/\pi)^2}{m}\sum_{i=2}^m\gamma_i^2=2(1-2/\pi)^2,
\eeq
\[
\frac{1}{m^2}\sum_{i=2}^m\text{var}(\alpha_i^2)\le \frac{1}{m}\max_iE( \alpha_i^4)=O(1/m),
\]
and using Lemma \ref{alpha.mom}
\[
\frac{1}{m^2}\sum_{2\le i\ne j\le m}\text{cov}(\alpha_i^2,\alpha_j^2)=\frac{m-1}{m}\text{cov}(\alpha_1^2,\alpha_2^2)=o(1),
\]
it follows that
\[
\sup_{t\in \R} \bigg|P\bigg(\frac{ \theta^\top \alpha}{m^{-1/2}\sigma_{sum}}\le t\bigg) -\Phi(t)\bigg|=o(1).
\]
Note that (\ref{conv.sigma.m}) holds, by Slutsky's theorem, we have
\beq \label{omega0.norm}
\frac{\Omega_0}{2^{1/2}(1-2/\pi)} \to_d N(0,1).
\eeq

\bigskip
\noindent{\bf Asymptotic normality of $A_n/n$.} We need the following results obtained by \cite{erdHos2012rigidity}.

\bel[Rigidity of Eigenvalues]
For (generalized) Wigner matrices, if $\gamma_j$ is the classical location of the $j$-th eigenvalue under the semicircle law ordered in increasing order, then the scaled $j$-th eigenvalue $\lambda_j/n^{1/2}$ is close to $\gamma_j$ in the sense that for some positive constants $C,c$
\[
\textup{pr}\bigg[\exists j: |\lambda_j/n^{1/2}-\gamma_j|\ge \frac{(\log n)^{c\log\log n}}{\{\min(j,n-j+1)\}^{1/3}n^{2/3}}\bigg]\le C\exp\{-(\log n)^{c\log\log n}\}
\]
for sufficiently large $n$.
\eel

As a consequence, for any $j=1,...,n$, we have 
\[
|\lambda_j/n^{1/2}-\gamma_j|=o_P(\{\min(j,n-j+1)\}^{-1/3}n^{-2/3+\delta})
\]
for any small $\delta>0$. In other words, the eigenvalue is near its classical location with an error of at most $N^{-1}(\log n)^{C\log\log n}$ for generalized Wigner matrices in the bulk and the estimate deteriorates by a factor $(n/j)^{1/3}$ near the edge $j\ll n$.
As a consequence, for any sufficiently small $\delta>0$,
\begin{align*}
|A_n/n-\Omega_0| &\le n^{-1/2}\sum_{i=2}^n|\lambda_i(W)/n^{1/2}-\gamma_i|\{\text{sgn}(u_1^\top)u_i\}^2= n^{-1/2}\sum_{i=2}^n\{\text{sgn}(u_1^\top)u_i\}^2\cdot o_P(n^{-2/3+\delta})
\end{align*}
Note that $\{\text{sgn}(u_1^\top)u_i\}^2=O_P(1)$, we have $|A_n/n-\Omega_0| \to 0$ in probability. So the asymptotic normality of $\Omega$ follows from Slutsky's theorem and (\ref{omega0.norm}).

\bigskip
\noindent{\bf Part III.} 
By definition, we have
\begin{align*}
\text{cov}(A_n/n, n^{-5/6}(B_n-2n^{1/2}\|u_1\|_1^2))&=\text{cov}\bigg( n^{-1}\sum_{i=2}^n \lambda_i \{\text{sgn}(u_1^\top)u_i\}^2, \frac{(\lambda_1-2{n^{1/2}})\|u_1\|_1^2}{n^{5/6}}\bigg).
\end{align*}
It suffices to control $\text{cov}\big( {\lambda_i} \{\text{sgn}(u_1^\top)u_i\}^2, n^{-1/3}{(\lambda_1/{n^{1/2}}-2)\|u_1\|_1^2}\big)$ for any $i=2,...,n$. Now we define 
\begin{align*}
\text{cov}\bigg( \lambda_i \{\text{sgn}(u_1^\top)u_i\}^2, n^{-1/3}{(\lambda_1/{n^{1/2}}-2)\|u_1\|_1^2}\bigg)=\text{cov}\big(\gamma_i(\text{sgn}(u_1^\top)u_i)^2,{n^{1/6}(\lambda_1/{n^{1/2}}-2)\|u_1\|_1^2}\big)+\mathcal{E}.
\end{align*}
\bel \label{lem.cov}
Under the conditions of Theorem 1, for any small constant $\epsilon>0$, it holds that $|\mathcal{E}|=o(n^{-1/6+2\epsilon})$ and $\textup{cov}\big(\gamma_i (\textup{sgn}(u_1^\top)u_i)^2,{n^{1/6}(\lambda_1/{n^{1/2}}-2)\|u_1\|_1^2}\big)=O(n^{-1/2+\epsilon})$.
\eel
Applying Lemma \ref{lem.cov} to the above equation, we complete the third part of our proof.

\qed

\section*{Supplementary materials}

Supplementary material includes the proofs of other theorems and the technical lemmas, as well as some supplementary tables and figures.

\bibliographystyle{chicago}
\bibliography{reference}

\newpage

\title{Supplement to ````The Asymptotic Distribution of Modularity in Weighted Signed Networks"}
\author{Rong Ma and Ian Barnett \\
	Department of Biostatistics, Epidemiology and Informatics\\
	University of Pennsylvania\\
	Philadelphia, PA 19104}
\date{}
\maketitle
\thispagestyle{empty}

\setcounter{section}{0}

\section{Proofs of Technical Lemmas}

\setcounter{lemma}{5}
\paragraph{Proof of Lemma 1.}
Since $(u_1,...,u_n)$ has a Haar measure on orthogonal group $O(n)$, following \cite{meckes_concentrationof}, $(u_1,...,u_n)$ has the same distribution as $(v_1,...,v_n)$, which is constructed as follows. Suppose $w_1,...,w_n$ are i.i.d. Gaussian vectors from $N(0,I_n)$. We define
\begin{align*}
&e_1 = w_1,   &v_1=\frac{e_1}{\|e_1\|_2}\\
&e_2 =w_2-\frac{\langle e_1,w_2\rangle}{\langle e_1,e_1 \rangle}e_1, &v_2=\frac{e_2}{\|e_2\|_2}\\
&...\\
&e_n =w_n-\sum_{k=1}^{n-1}\frac{\langle e_k,w_n\rangle}{\langle e_k,e_k \rangle}e_k, &v_n=\frac{e_n}{\|e_n\|_2}.
\end{align*} 
Now since $\{u_1,...,u_n\}$ are exchangeable, the distribution of $\text{sgn}(u_{1}^\top)u_i$ for $i\ne 1$ is the same as the distribution of $\text{sgn}(v_{2}^\top)v_1$. Thus it suffices to consider the following problem.
Consider $X=(X_1,...,X_n)$ where $X_i\sim N(0,1)$, and $Y=(Y_1,...,Y_n)$ where $Y_i\sim N(0,1)$. $X$ and $Y$ are independent. Let $Z$ defined by
\[
W = Y-\frac{\langle X,Y\rangle}{\langle X,X\rangle}X,\quad\quad Z=W/\|W\|_2
\]
be the Gram-Schmitt transformed $Y$ so that $\langle X,Z\rangle=0$. It suffices to prove
\[
\Delta= \sum_{i=1}^n \text{sgn}(Z_i)X_i/\|X\|_2= \sum_{i=1}^n \text{sgn}(W_i)X_i/\|X\|_2
\]
is asymptotically normal.
Let $A_i=\text{sgn}(W_i)X_i$ and
\[
\Delta =  \sum_{i=1}^n \text{sgn}(W_i)X_i/\|X\|_2=\frac{1}{\|X\|_2}\sum_{i=1}^nA_i.
\]
By definition, 
\[
A_i = \text{sgn}\bigg( Y_i-\frac{\langle X,Y\rangle}{\|X\|_2^2} X_i\bigg)X_i.
\]
We first claim that, conditional on $X$, the random variable $\text{sgn}\big( Y_i-\frac{\langle X,Y\rangle}{\|X\|_2^2} X_i\big)$ is a Bernoulli random variable taking values from $\{-1,1\}$ with even probability. To see this, notice that 
\[
P\bigg(Y_i-\frac{\langle X,Y\rangle}{\|X\|_2^2} X_i>0\bigg|X\bigg)=P\bigg(-Y_i-\frac{\langle X,-Y\rangle}{\|X\|_2^2} X_i>0\bigg|X\bigg)=P\bigg(Y_i-\frac{\langle X,Y\rangle}{\|X\|_2^2} X_i<0\bigg|X\bigg),
\]
as the distribution of $Y$ is the same as the distribution of $-Y$. It then follows that $A_i|X$ is a Bernoulli random variable taking values from $\{-X_i,X_i\}$ with even probability. As a result, the density of $A_i$ can be calculated by integration over the marginal distribution of $X$, which leads to a standard normal density. Hence,
\beq \label{A_i.eq}
A_i\sim N(0,1).
\eeq
To obtain the covariance between $A_i,A_j$ for $i\ne j$, note that
\begin{align*}
\text{Cov}(A_i,A_j)&=\E A_iA_j=\E\text{sgn}\bigg( Y_i-\frac{\langle X,Y\rangle}{\|X\|_2^2} X_i\bigg)\text{sgn}\bigg( Y_j-\frac{\langle X,Y\rangle}{\|X\|_2^2} X_j\bigg)X_iX_j\\
&= \E\bigg[X_iX_j \E \bigg( \text{sgn}\bigg( Y_i-\frac{\langle X,Y\rangle}{\|X\|_2^2} X_i \bigg)\text{sgn}\bigg( Y_j-\frac{\langle X,Y\rangle}{\|X\|_2^2} X_j \bigg) \bigg| X \bigg) \bigg].
\end{align*}
We can write
\[
\E \bigg( \text{sgn}\bigg( Y_i-\frac{\langle X,Y\rangle}{\|X\|_2^2} X_i \bigg)\text{sgn}\bigg( Y_j-\frac{\langle X,Y\rangle}{\|X\|_2^2} X_j \bigg) \bigg| X \bigg)=2p_{ij}-1
\]
where
\begin{align}
p_{ij}= P\bigg( \bigg[ Y_i-\frac{\langle X,Y\rangle}{\|X\|_2^2} X_i \bigg]\bigg[ Y_j-\frac{\langle X,Y\rangle}{\|X\|_2^2} X_j \bigg] \ge 0 \bigg| X \bigg).
\end{align}
Inside the probability measure, we have a quadratic form of $Y_i$'s. Specifically, define
\[
y = (Y_1,Y_2,..., Y_n)^\top,\quad\quad
\ell_i = \bigg( -\frac{X_1X_i}{\|X\|_2^2},-\frac{X_2X_i}{\|X\|_2^2},...,1-\frac{X_i^2}{\|X\|_2^2},...,-\frac{X_nX_i}{\|X\|_2^2}\bigg)^\top,
\]
we have for $i\ne j$,
\begin{align*}
\bigg[ Y_i-\frac{\langle X,Y\rangle}{\|X\|_2^2} X_i \bigg]\bigg[ Y_j-\frac{\langle X,Y\rangle}{\|X\|_2^2} X_j \bigg]=y^\top \ell_i\cdot\ell_j^\top y.
\end{align*}
For fixed $\ell_i$ and $\ell_j$, we have $y^\top \ell_i\sim N(0,\|\ell_i\|_2^2)$ and  $y^\top \ell_j\sim N(0,\|\ell_j\|_2^2)$. Then
\[
(y^\top \ell_i,y^\top \ell_j)\sim N(\bf{0}, A),\quad\quad A=\begin{bmatrix}
\|\ell_i\|_2^2 & \ell_i^\top\ell_j \\
\ell_i^\top\ell_j  & \|\ell_j\|_2^2
\end{bmatrix}.
\]
Hence
\beq
P(y^\top \ell_i\cdot y^\top \ell_j\ge 0 |\ell_i,\ell_j)=\frac{2}{\sqrt{1-\rho^2}}\int_{-\infty}^0\phi(x)\int_{-\infty}^0\phi\bigg(\frac{x-\rho y}{\sqrt{1-\rho^2}} \bigg)dxdy=\frac{1}{\pi}\arcsin \rho+\frac{1}{2}
\eeq
where 
\begin{align*}
\rho &= \frac{\ell_i^\top\ell_j}{\|\ell_i\|_2\|\ell_j\|_2}= -\frac{X_iX_j}{\|X\|_2^2}\bigg(1-\frac{X_i^2}{\|X\|_2^2}\bigg)^{-1/2}\bigg(1-\frac{X_j^2}{\|X\|_2^2}\bigg)^{-1/2}\\
&=-\frac{X_iX_j}{\sqrt{\sum_{k\ne i}X_k^2}\sqrt{\sum_{k\ne j}X_k^2}}.
\end{align*}
Thus
\beq
p_{ij}=\frac{1}{\pi}\arcsin \rho+\frac{1}{2},
\eeq
and hence
\[
\text{Cov}(A_i,A_j) = \E[X_iX_j(2p_{ij}-1)]= \frac{2}{\pi} \E[X_iX_j\arcsin \rho].
\]
Now since $\rho\to 0$ in probability as $n\to \infty$ and
\[
\arcsin \rho = \rho+O(\rho^3),
\]
we have
\[
\bigg|\text{Cov}(A_i,A_j) - \frac{2}{\pi} \E[ X_iX_j\rho]\bigg| \le \sqrt{\E \rho^6}= O(1/n^3).
\]
where we used Cauchy-Schwartz in the first inequality. Calculate that
\[
\frac{2}{\pi} \E[ X_iX_j\rho]=-\frac{2}{\pi}\E\bigg[ \frac{X_i^2X_j^2}{\sqrt{\sum_{k\ne i}X_k^2}\sqrt{\sum_{k\ne j}X_k^2}}\bigg]
\]
It follows that
\beq \label{cov.eq}
\text{Cov}(A_i,A_j)=-\frac{2}{n\pi}+o(1/n).
\eeq
Combining (\ref{A_i.eq}) and (\ref{cov.eq}), we have
\beq \label{norm.A_i}
\frac{1}{\sqrt{n}}\sum_{i=1}^n A_i \sim N(0,\sigma^2)
\eeq
where
\[
\sigma^2=\text{Var}(A_i)+(n-1)\text{Cov}(A_i,A_j)=1-\frac{2}{\pi}+o(1).
\]
On the other hand, by concentration inequality for independent sub-exponential random variables
\[
P\bigg( \bigg|\frac{\|X\|_2^2}{n}-1\bigg|\ge \sqrt{\frac{\log n}{n}}\bigg)\le \frac{1}{n^c}.
\]
Then using the inequality $(\sqrt{a}-\sqrt{b})^2\le |a-b|$, we have, with probability at least $1-O(n^{-c})$ for some $c>0$,
\[
\bigg|\frac{\|X\|_2}{\sqrt{n}}-1\bigg|\le \bigg|\frac{\|X\|_2^2}{n}-1\bigg|\le \sqrt{\frac{\log n}{n}}.
\]
Therefore, with probability at least $1-O(n^{-c}),$
\beq \label{error}
\bigg|\Delta-\frac{1}{\sqrt{n}}\sum_{i=1}^n A_i\bigg| \le \bigg|\frac{\sqrt{n}}{\|X\|_2}-1  \bigg|\cdot\bigg|\frac{1}{\sqrt{n}}\sum_{i=1}^n A_i\bigg|\le  c\frac{\log n}{\sqrt{n}},
\eeq
and we can write
\[
\Delta = \frac{1}{\sqrt{n}}\sum_{i=1}^n A_i+O_P(n^{-1/2}\log n).
\]
Along with (\ref{norm.A_i}), we have
\[
\Delta \to_d N(0,1-2/\pi).
\]
\qed


\paragraph{Proof of Lemma 2.}	
Recall the previous characterization of the Haar measure on $O(n)$, we set
\[
e_1=w_1,\quad e_2=w_2-\frac{\langle e_1,w_2\rangle}{\langle e_1,e_1\rangle}e_1,
\]
\beq \label{haar.char}
e_3=w_3-\frac{\langle e_1,w_3\rangle}{\langle e_1,e_1\rangle}e_1-\frac{\langle e_2,w_3\rangle}{\langle e_2,e_2\rangle}e_2,
\eeq
for i.i.d. $w_1,w_2,w_3 \sim N(0,I_n)$
and set
\beq \label{1.9}
u_i=\frac{e_1}{\|e_1\|_2}, \quad u_1=\frac{e_2}{\|e_2\|_2},\quad u_j=\frac{e_3}{\|e_3\|_2}.
\eeq
It follows that
\begin{align*}
\E [\text{sgn}(u_1^\top)u_i \text{sgn}(u_1^\top)u_j]
&=\E\bigg[\sum_{i=1}^n\text{sgn}(e_{2i})\frac{w_{1i}}{\|w_1\|_2}\bigg]\bigg[\sum_{i=1}^n \text{sgn}(e_{2i})\frac{e_{3i}}{\|e_3\|_2}\bigg]\\
&=\sum_{i=1}^n \E\bigg[ \frac{w_{1i}}{\|w_1\|_2}\frac{e_{3i}}{\|e_3\|_2}\bigg]+\sum_{i\ne j}\E\bigg[\text{sgn}(e_{2i})\text{sgn}(e_{2j})\frac{w_{1i}}{\|w_1\|_2}\frac{e_{3j}}{\|e_3\|_2}\bigg]\\
&=I_1+I_2.
\end{align*}
For the first term $I_1$, note that
\begin{align*}
\E\frac{w_{1i}}{\|w_1\|_2}\frac{e_{3i}}{\|e_3\|_2}=\E\bigg[\frac{w_{1i}}{\|w_1\|_2} \E\bigg[\frac{e_{3i}}{\|e_3\|_2}  \bigg| w_1,w_2\bigg] \bigg].
\end{align*}
Apparently, due to symmetry with respect to $w_3$, $\E [e_{3i}/\|e_3\|_2|w_1,w_2]=0$. 
Thus $I_1=0$.
For the second term $I_2$, we have
\begin{align*}
\E\bigg[\text{sgn}(e_{2i})\text{sgn}(e_{2j})\frac{w_{1i}}{\|w_1\|_2}\frac{e_{3j}}{\|e_3\|_2}\bigg]=\E\bigg[\E\bigg[\frac{e_{3j}}{\|e_3\|_2}\bigg|w_1,w_2\bigg]\text{sgn}(e_{2i})\text{sgn}(e_{2j})\frac{w_{1i}}{\|w_1\|_2}\bigg]=0
\end{align*}
To prove the second statement, we show that $(\textup{sgn}(u_1^\top)u_i,\textup{sgn}(u_1^\top)u_j)$ has an asymptotically bivariate normal distribution. Toward this end, we show that for any $\nu=(\nu_1,\nu_2) \in \R^2$, it holds that
\beq \label{joint.norm}
\nu_1\textup{sgn}(u_1^\top)u_i+\nu_2\textup{sgn}(u_1^\top)u_j \to_d N(0,\|\nu\|^2_2)
\eeq
for some $\sigma^2(\nu)\ge 0$. Again, using the Gaussian representation in (\ref{haar.char}), we define
\beq
u_i=\frac{e_1}{\|e_1\|_2}, \quad u_j=\frac{e_2}{\|e_2\|_2},\quad u_1=\frac{e_3}{\|e_3\|_2}.
\eeq
It follows that
\begin{align*}
&\nu_1\textup{sgn}(u_1^\top)u_i+\nu_2\textup{sgn}(u_1^\top)u_j \\
&=\sum_{k=1}^n \textup{sgn}\bigg(w_{3k}-\frac{\langle e_1,w_3\rangle}{\langle e_1,e_1\rangle}e_{1k}-\frac{\langle e_2,w_3\rangle}{\langle e_2,e_2\rangle}e_{2k} \bigg)\bigg(\nu_1\frac{ w_{1k}}{\|w_{1}\|_2}-\frac{\nu_2}{\|e_2\|_2}\bigg(w_{2k}-\frac{\langle w_1,w_2\rangle}{\| w_1\|_2^2}w_{1k}\bigg) \bigg)\\
&=\frac{1}{\sqrt{n}}\sum_{k=1}^n \textup{sgn}\bigg(w_{3k}-\frac{\langle e_1,w_3\rangle}{\langle e_1,e_1\rangle}e_{1k}-\frac{\langle e_2,w_3\rangle}{\langle e_2,e_2\rangle}e_{2k} \bigg)(\nu_1w_{1k}-\nu_2w_{2k})+Rem.
\end{align*}
In the following, we show that 
\beq \label{bi.norm}
\frac{1}{\sqrt{n}}\sum_{k=1}^n \textup{sgn}\bigg(w_{3k}-\frac{\langle e_1,w_3\rangle}{\langle e_1,e_1\rangle}e_{1k}-\frac{\langle e_2,w_3\rangle}{\langle e_2,e_2\rangle}e_{2k} \bigg)(\nu_1w_{1k}-\nu_2w_{2k})\to_d N(0,\sigma^2(\nu)),\quad Rem=o_P(1).
\eeq
On the one hand, using the same argument as in the proof of Lemma 1, we have, the random variable
\[
\textup{sgn}\bigg(w_{3k}-\frac{\langle e_1,w_3\rangle}{\langle e_1,e_1\rangle}e_{1k}-\frac{\langle e_2,w_3\rangle}{\langle e_2,e_2\rangle}e_{2k} \bigg)(\nu_1w_{1k}-\nu_2w_{2k}) \bigg| w_1,w_2
\]
is a Bernoulli random variable taking values in $\{\nu_1w_{1k}-\nu_2w_{2k},-\nu_1w_{1k}+\nu_2w_{2k}\}$ with even probability. By joint normality of $w_1$ and $w_2$, one can obtain 
\beq \label{3.norm}
\textup{sgn}\bigg(w_{3k}-\frac{\langle e_1,w_3\rangle}{\langle e_1,e_1\rangle}e_{1k}-\frac{\langle e_2,w_3\rangle}{\langle e_2,e_2\rangle}e_{2k} \bigg)(\nu_1w_{1k}-\nu_2w_{2k}) \sim N(0,\|\nu\|_2^2),
\eeq
for each $1\le k\le n$.
Thus the first statement of (\ref{bi.norm}) holds. To show $R=o_P(1)$,  note that
\begin{align*}
Rem&=\frac{1}{\sqrt{n}}\sum_{k=1}^n \textup{sgn}(e_{3k})\bigg( (1-\sqrt{n}/\|e_2\|_2)\nu_2w_{2k}+(1-\sqrt{n}/\|w_1\|_2)\nu_1w_{1k}+\frac{\sqrt{n}}{\|e_2\|_2}\frac{w_1^\top w_2}{\|w_1\|_2^2}\nu_2w_{1k} \bigg)\\
&\le \nu_2\bigg|\frac{1}{\sqrt{n}}\sum_{k=1}^n \textup{sgn}(e_{3k})w_{2k}\bigg|\cdot\bigg|1-\frac{\sqrt{n}}{\|e_2\|_2}  \bigg|+\nu_1\bigg|\frac{1}{\sqrt{n}}\sum_{k=1}^n \textup{sgn}(e_{3k})w_{1k}\bigg|\cdot\bigg|1-\frac{\sqrt{n}}{\|w_1\|_2}  \bigg|\\
&\quad+\nu_2\bigg|\frac{1}{\sqrt{n}}\sum_{k=1}^n \textup{sgn}(e_{3k})w_{1k}\bigg|\cdot\bigg| \frac{\sqrt{n}}{\|e_2\|_2}\frac{w_1^\top w_2}{\|w_1\|_2^2}\bigg|.
\end{align*}
By concentration inequality for sub-exponential random variables, we have
\[
\bigg|1-\frac{\sqrt{n}}{\|w_\ell\|_2} \bigg|=O_P(n^{-1/2}),\quad \ell=1,2,
\]
and
\[
|w_1^\top w_2|=O_P(\sqrt{n}).
\]
In addition, since $ \|e_2\|_2^2= \|w_2\|_2^2-|w_1^\top w_2|^2/\|w_1\|_2^2$, we have
\[
\bigg|1-\frac{\sqrt{n}}{\|e_2\|_2} \bigg|\le C\bigg|1-\frac{\|e_2\|_2^2}{n} \bigg|\le C_1\bigg|1-\frac{\|w_2\|_2^2}{n} \bigg|+C_2\bigg| \frac{|w_1^\top w_2|^2}{n\|w_1\|_2^2} \bigg|=O_P(n^{-1/2}).
\]
Using the same conditional argument that leads to (\ref{3.norm}), we also have
\[
\frac{1}{\sqrt{n}}\sum_{k=1}^n \textup{sgn}(e_{3k})w_{\ell k}\sim N(0,1),\qquad \ell=1,2.
\]
As a result, we obtain $Rem=o_P(1)$, which completes the proof of (\ref{joint.norm}). By the Cram\'er-Wold theorem, $(\textup{sgn}(u_1^\top)u_i,\textup{sgn}(u_1^\top)u_j) $ is asymptotically bivariate normal. Since we just proved
\[
\E [\textup{sgn}(u_1^\top)u_i\textup{sgn}(u_1^\top)u_j]=0
\]
for all $n\ge 1$ and all $i,j\in\{2,...,n\}$, we have	
\[
(\textup{sgn}(u_1^\top)u_i,\textup{sgn}(u_1^\top)u_j) \to_d N(0,\mathbf{B}),\quad\quad 
\mathbf{B} = \begin{bmatrix}
1-2/\pi & 0\\
0 & 1-2/\pi
\end{bmatrix}.
\]
Now, in order to obtain the second statement of Lemma 2, we need to establish the convergence of moments from the convergence in distribution using the following lemma, which can be find in many standard texts such as Theorem 1.8 in \cite{shao2003}.

\bel \label{shao.lem}
Let $X,X_1,X_2,...$ be random $k$-vectors. Suppose that $X_n\to_d X$. Then for any $r>0$, $\lim_{n\to\infty}\E \|X_n\|_r^r=\E \|X\|_r^r<\infty$ if and only if $\{ \|X_n\|_r^r \}$ is uniformly integrable. In particular, a sufficient condition for uniform integrability of $\{ \|X_n\|_r^r \}$ is that $\sup_n \E \|X_n\|_r^{r+\delta}<\infty$ for a $\delta>0$.
\eel

It suffices to check $\sup_n\E (\textup{sgn}(u_1^\top)u_i)^r(\textup{sgn}(u_1^\top)u_j)^r<\infty$
for some $r\ge 3$. To see this, for any $n>0$, by Cauchy-Schwartz inequality and the first inequality in (\ref{error}),
\begin{align*}
\E (\textup{sgn}(u_1^\top)u_i)^r(\textup{sgn}(u_1^\top)u_j)^r &\le \sqrt{\E (\textup{sgn}(u_1^\top)u_i)^{2r}}\sqrt{\E (\textup{sgn}(u_1^\top)u_j)^{2r} }\\
&\le  \E \bigg(\mathcal{N}+ \mathcal{N}\bigg|\frac{\sqrt{n}}{\|X\|_2}-1  \bigg|\bigg)^{2r} \\
&\le 4^r (\E \mathcal{N}^{2r}+ \E \bigg|\frac{\sqrt{n}}{\|X\|_2}-1  \bigg|^{2r}\cdot\mathcal{N}^{2r})\\
&<C_r,
\end{align*}
where $\mathcal{N}=\frac{1}{\sqrt{n}}\sum_{i=1}^n A_i$ using the notation of (\ref{error}). $C_r<\infty$ is some constant only depending on $r$, and the last inequality follows from $\|X\|_2^2/n \to 1$ a.s. (strong law of large numbers) and the normality of $\mathcal{N}$. This completes the proof of the uniform integrability of $\{(\textup{sgn}(u_1^\top)u_i)^2(\textup{sgn}(u_1^\top)u_j)^2 \}_{n\ge 1}$. 
\qed

%

\paragraph{Proof of Lemma 3.}
The proof follows essentially the proof of Theorem 1 in \cite{meckes2007central}.
Let $I$ be chosen uniformly from $\{1,...,n\}$, independently of $X$, and define
\[
X' = X-2X_{I}e_I.
\]
Then $(X,X')$ is an exchangeable pair of random vectors by assumption.
We need the following lemma proved by \cite{stein1986approximate}.
\bel \label{stein.lem}
Let $(W,W')$ be an exchangeable pair of random variables such that
\[
\E W=0,\quad \E W^2=1,
\]
and 
\[
\E [W-W'|W]=\lambda W
\]
for some $\lambda \in (0,1)$. Then 
\[
\sup_{t\in \R}|P(W\le t)-\Phi(t)|\le \frac{1}{\lambda}\sqrt{\text{Var} \E[(W-W')^2|W]}+(2\pi)^{-1/4}\sqrt{\frac{1}{\lambda}\E |W-W'|^3}.
\]
\eel

Define $\sigma^2_\theta=\sum_{i=1}^n\theta_i^2\sigma_i^2$, $W=W_\theta=\frac{\langle X,\theta\rangle}{\sigma_\theta}$ and $W'=\frac{\langle X',\theta \rangle}{\sigma_\theta}$. Now $\E W=0$ and $\E W^2=1$ since $X$ is isotropic, and 
\begin{align*}
\E[W-W'|W] = \E \bigg[\frac{1}{n\sigma_\theta}\sum_{i=1}^n 2X_{i}\theta_{i}\bigg| W \bigg]=\frac{2}{n\sigma_\theta}\E\bigg[ \bigg\langle \sum_{i=1}^n e_ie_i^\top X,\theta \bigg\rangle \bigg| W\bigg]=\frac{2W}{n}.
\end{align*}
To apply Lemma \ref{stein.lem}, it remains to estimate the quantities
\[
\text{Var} \E[(W-W')^2|W],\quad \E |W-W'|^3.
\]
Firstly,
\[
\E(\E[(W-W')^2|W]) = \E(\E[W^2+(W')^2-2WW'|W])=\frac{4}{n},
\]
and y the conditional form of Jensen's inequality
\begin{align*}
\E(\E[(W-W')^2|W])^2 \le \E(\E[(W-W')^2|X])^2=\frac{1}{\sigma^4_\theta}\E(\E[(2X_I\theta_{I})^2|X])^2=\frac{16}{\sigma^4_\theta n^2}\sum_{i,j=1}^n\theta_i^2\theta_j^2\E[X_i^2X_j^2],
\end{align*}
so
\begin{align*}
\text{Var}\E[(W-W')^2|W]&=\E(\E[(W-W')^2|W])^2-\frac{16}{n^2}\\
&\le \frac{16}{n^2}\bigg(\frac{1}{\sigma_\theta^4} \sum_{i,j=1}^n\theta_i^2\theta_j^2\E[X_i^2X_j^2]-1\bigg).
\end{align*}
Next, we calculate that
\begin{align*}
\E|W-W'|^3=\frac{8}{\sigma_\theta^3}\E|X_I\theta_I|^3=\frac{8}{\sigma_\theta^3 n}\sum_{i=1}^n|\theta_i|^3\E|X_i|^3\le \frac{8}{\sigma_\theta^3n}\bigg(\max_{i}\E|X_i|^3\bigg)\sum_{i=1}^n|\theta_i|^3.
\end{align*}
The proof is complete by inserting these estimates into Lemma \ref{stein.lem}.
\qed


\paragraph{Proof of Lemma 4.}
For the first statement, by definition 
\begin{align*}
\E \alpha_i^2 &= \frac{1}{2}\E[\zeta_i^4+\eta_i^4-2\zeta_i^2\eta_i^2]\\
&=\frac{\gamma_i^2}{2}\big(\E[\text{sgn}(u_1^\top)u_i]^4+\E[\text{sgn}(u_1^\top)u_j]^4 -2\E[\text{sgn}(u_1^\top)u_i]^2[\text{sgn}(u_1^\top)u_j]^2 \big).
\end{align*}
By Lemma 2, the last term in the last expression is
\[
2\E[\text{sgn}(u_1^\top)u_i^\top]^2[\text{sgn}(u_1^\top)u_j^\top]^2=2(1-\pi/2)^2+o(1).
\]
On the other hand, by asymptotic normality of $\text{sgn}(u_1^\top)u_j$ for each $j=2,...,n$ and Lemma 6, we have
\[
\E[\text{sgn}(u_1^\top)u_j]^4=3(1-2/\pi)^2+o(1).
\]
It then follows that
\beq \label{unif}
\E\alpha_i^2=2\gamma_i^2(1-2/\pi)^2+o(1).
\eeq
The second statement follows from Lemma 1 and Lemma 2. Specifically, using the same argument that leads to (\ref{bi.norm}), one can show that for any fixed $i,j\in \{2,...,m\}, i\ne j$, 
\[
(\textup{sgn}(u_1^\top)u_i,\textup{sgn}(u_1^\top)u_j,\textup{sgn}(u_1^\top)u_{n-i+1},\textup{sgn}(u_1^\top)u_{n-j+1}) \to_d N(0, (1-2/\pi){\bf I}_4).
\] 
As a result, the random variables $\{\zeta_i,\eta_i,\zeta_j,\eta_j\}$ are also asymptotically independent Gaussian random variables, and so does statistics $\alpha_i$ and $\alpha_j$. The second statement then follows again from Lemma \ref{shao.lem} and (\ref{unif}) which guarantees the uniform integrability conditions.
\qed

\paragraph{Proof of Lemma 6.}
By definition, we have
\begin{align*}
|\mathcal{E}|&=\bigg|\text{Cov}\bigg((\lambda_i/\sqrt{n}-\gamma_i)(\text{sgn}(u_1^\top)u_i)^2,n^{-1/2}\|u_1\|_1^2(n^{2/3}(\lambda_1/\sqrt{n}-2))  \bigg)  \bigg| \\
&\le \sqrt{\Var((\lambda_i/\sqrt{n}-\gamma_i)(\text{sgn}(u_1^\top)u_i)^2)}\sqrt{\Var(n^{-1/2}\|u_1\|_1^2(n^{2/3}(\lambda_1/\sqrt{n}-2)) )}\\
&\le {[\E(\lambda_i/\sqrt{n}-\gamma_i)^4]^{1/4}[ \E(\text{sgn}(u_1^\top)u_i)^8]^{1/4}}{[\E n^{-2}\|u_1\|_1^8]^{1/4}[\E n^{8/3}(\lambda_1/\sqrt{n}-2)^4]^{1/4} },
\end{align*}
by using  Cauchy-Schwartz inequality.
From Lemma 1 and Lemma 6, we have $[ \E(\text{sgn}(u_1^\top)u_i)^8]^{1/4}=O(1)$. From the Gaussian representation of $u_1$ defined in (A2) of the main paper, we have $[\E n^{-2}\|u_1\|_1^8]^{1/4}=O(\sqrt{n})$. From Lemma 5 (rigidity of eigenvalues) and the almost sure bound $\lambda_1= (2+o(1))\sqrt{n}$ given by \cite{bai1988necessary}, we have for any sufficiently small $\epsilon>0$,
\[
[\E(\lambda_i/\sqrt{n}-\gamma_i)^4]^{1/4}=o(n^{-2/3+\epsilon}).
\]
Combining these bounds, we have $|\mathcal{E}|=o(n^{-1/6+2\epsilon})$. Now recall that the eigenvalues and the eigenvectors of GOE are independent. It holds that
\begin{align}
\bigg|\text{Cov}\bigg(\gamma_i (\text{sgn}(u_1^\top)u_i)^2,\frac{n^{2/3}(\lambda_1/\sqrt{n}-2)\|u_1\|_1^2}{\sqrt{n}}\bigg)\bigg|&=\bigg|\E n^{2/3}(\lambda_1/\sqrt{n}-2) \cdot \text{Cov}\bigg(\gamma_i (\text{sgn}(u_1^\top)u_i)^2,\frac{\|u_1\|_1^2}{\sqrt{n}}\bigg)\bigg| \nonumber \\
&=O(n^\epsilon)\cdot \bigg|\text{Cov}\bigg(\gamma_i (\text{sgn}(u_1^\top)u_i)^2,\sqrt{n}\big(\frac{\|u_1\|_1^2}{{n}}-\frac{\pi}{2}\big)\bigg)\bigg|. \label{cov}
\end{align}
Now we analyse the above covariance more carefully. Firstly, using the Gaussian representation (\ref{haar.char}) with
\[
u_1=\frac{e_1}{\|e_1\|_2},\quad u_i=\frac{e_2}{\|e_2\|_2},
\]
direct calculation yields
\begin{align*}
\sqrt{n}\bigg(\frac{\|u_1\|_1^2}{{n}}-\frac{\pi}{2}\bigg) &=\frac{1}{\sqrt{n}}+\frac{1}{\sqrt{n}}\bigg( \frac{1}{n}\sum_{i\ne j} |w_{1i}w_{1j}|-\frac{2n}{\pi} \bigg)+\frac{2}{\pi}\sqrt{n}\bigg(\frac{\|w_{1}\|_2^2}{n}-1 \bigg)\\
&\quad+\bigg(\frac{2}{\pi}-\frac{\sum_{i\ne j}|w_{1i}w_{1j}|}{n^2(1+(\|w_1\|_2^2/n-1))}\bigg)\sqrt{n}\bigg(\frac{\|w_1\|_2^2}{n}-1 \bigg)\\
&\equiv\frac{2}{\pi}\sqrt{n}\bigg(\frac{\|w_{1}\|_2^2}{n}-1 \bigg)+\frac{1}{\sqrt{n}}\Delta_1,
\end{align*}
and
\begin{align*}
\text{sgn}(u_1^\top)u_i&= \frac{1}{\sqrt{n}}\sum_{j=1}^n\text{sgn}(w_{1j})w_{2j}+ \frac{1}{\sqrt{n}}\bigg(\frac{\sqrt{n}}{\|e_2\|_2}- 1 \bigg)\sum_{j=1}^n\text{sgn}(w_{1j})w_{2j} - \frac{w_1^\top w_2}{\|w_1\|_2^2\|e_2\|_2}\sum_{j=1}^n|w_{1j}|,
\end{align*}
or
\[
(\text{sgn}(u_1^\top)u_i)^2\equiv\bigg(\frac{1}{\sqrt{n}}\sum_{j=1}^n\text{sgn}(w_{1j})w_{2j}\bigg)^2+\frac{1}{\sqrt{n}}\Delta_2.
\]
Hence, we have
\begin{align*}
&\bigg|\text{Cov}\bigg(\gamma_i (\text{sgn}(u_1^\top)u_i)^2,\sqrt{n}\big(\frac{\|u_1\|_1^2}{{n}}-\frac{\pi}{2}\big)\bigg)\bigg| \\
&\lesssim \bigg| \text{Cov}\bigg( \sqrt{n}\bigg(\frac{\|w_{1}\|_2^2}{n}-1 \bigg),\bigg(\frac{1}{\sqrt{n}}\sum_{j=1}^n\text{sgn}(w_{1j})w_{2j}\bigg)^2 \bigg)\bigg|+\frac{1}{\sqrt{n}}|\text{Cov}(\sqrt{n}\bigg(\frac{\|w_{1}\|_2^2}{n}-1 \bigg), \Delta_2)|\\
&\quad +\frac{1}{\sqrt{n}}\bigg|\text{Cov} \bigg( \Delta_1, \bigg(\frac{1}{\sqrt{n}}\sum_{j=1}^n\text{sgn}(w_{1j})w_{2j}\bigg)^2\bigg)  \bigg|+\frac{1}{\sqrt{n}}|\text{Cov}(\Delta_1,\Delta_2)|.
\end{align*}
To show that the last three terms are all of order $O(n^{-1/2})$, we use Cauchy-Schwartz inequality. It suffices to control the variances or the second moments of $\Delta_1,\Delta_2, \sqrt{n}\big(\frac{\|w_{1}\|_2^2}{n}-1 \big)$ and $\big(\frac{1}{\sqrt{n}}\sum_{j=1}^n\text{sgn}(w_{1j})w_{2j}\big)^2$, respectively. Specifically, using Lemma A1 and Lemma 6, we have $\Var[\big(\frac{1}{\sqrt{n}}\sum_{j=1}^n\text{sgn}(w_{1j})w_{2j}\big)^2]=O(1)$. Using standard CLT and Lemma 6, we have $\Var[\sqrt{n}\big(\frac{\|w_{1}\|_2^2}{n}-1 \big)]=O(1)$. For $\Delta_1$, we can control its second moment by using Lemma 3 for the term $\big( \frac{1}{n}\sum_{i\ne j} |w_{1i}w_{1j}|-\frac{2n}{\pi} \big)$, strong law of large numbers for $\|w_1\|_2^2/n-1$, and Lemma 6, to obtain $\E \Delta_1^2=O(1)$. Lastly, one can similarly obtain $\E \Delta_2^2=O(1)$ using Lemma 1 for the term $\frac{1}{\sqrt{n}}\sum_{j=1}^n\text{sgn}(w_{1j})w_{1j}$ and $\sqrt{n}w_1^\top w_2$, strong law of large numbers for $\frac{1}{n}\sum_{j=1}^n|w_{1j}|$ and $\|w_1\|_2^2/n$, and Lemma 6. As a result, we have
\begin{align*}
&\bigg|\text{Cov}\bigg(\gamma_i (\text{sgn}(u_1^\top)u_i)^2,\sqrt{n}\big(\frac{\|u_1\|_1^2}{{n}}-\frac{\pi}{2}\big)\bigg)\bigg| \\
&\lesssim \bigg| \text{Cov}\bigg( \sqrt{n}\bigg(\frac{\|w_{1}\|_2^2}{n}-1 \bigg),\bigg(\frac{1}{\sqrt{n}}\sum_{j=1}^n\text{sgn}(w_{1j})w_{2j}\bigg)^2 \bigg)\bigg|+O(1/\sqrt{n}).
\end{align*}
Now since
\begin{align*}
&\text{Cov}\bigg( \sqrt{n}\bigg(\frac{\|w_{1}\|_2^2}{n}-1 \bigg),\bigg(\frac{1}{\sqrt{n}}\sum_{j=1}^n\text{sgn}(w_{1j})w_{2j}\bigg)^2 \bigg)\\
&=\frac{1}{n^{3/2}}\text{Cov}\bigg( {\sum_{j=1}^n w_{1j}},\sum_{1\le j, k\le n}\text{sgn}(w_{1j})\text{sgn}(w_{1k})w_{2j}w_{2k}\bigg)\\
&=\frac{1}{n^{3/2}}\sum_{1\le j,k,\ell\le n} \text{Cov}(w_{1\ell}, \text{sgn}(w_{1j})\text{sgn}(w_{1k})w_{2j}w_{2k})\\
&=0,
\end{align*}
where the last equation holds since 
\begin{align*}
\text{Cov}(w_{1\ell}, \text{sgn}(w_{1j})\text{sgn}(w_{1k})w_{2j}w_{2k})&=\E w_{1\ell}\text{sgn}(w_{1j})\text{sgn}(w_{1k})w_{2j}w_{2k}-\E w_{1\ell}\E \text{sgn}(w_{1j})\text{sgn}(w_{1k})w_{2j}w_{2k}\\
&=\E w_{1\ell}\text{sgn}(w_{1j})\text{sgn}(w_{1k})\E w_{2j}w_{2k}-\E w_{1\ell}\E \text{sgn}(w_{1j})\text{sgn}(w_{1k})\E w_{2j}w_{2k}\\
&=\E w_{2j}w_{2k} (\E w_{1\ell}\text{sgn}(w_{1j})\text{sgn}(w_{1k})-\E w_{1\ell}\E \text{sgn}(w_{1j})\text{sgn}(w_{1k}))\\
&=0,
\end{align*}
for any $1\le j,k,\ell\le n$. Back to (\ref{cov}), we have shown
\[
\bigg|\text{Cov}\bigg(\gamma_i (\text{sgn}(u_1^\top)u_i)^2,\frac{n^{2/3}(\lambda_1/\sqrt{n}-2)\|u_1\|_1^2}{\sqrt{n}}\bigg)\bigg|=O(n^{-1/2+\epsilon}),
\]
for some sufficiently small $\epsilon>0$. This completes the proof.
\qed

\section{Proof of Theorem 2}

Since $Q=A_n+B_n$ and $B_n=\lambda_1\|u_1\|_1^2$, we have
\begin{align*}
\text{Cov}(Q/n,n^{1/6}\lambda_1)&=\text{Cov}\bigg( \frac{A_n}{n},n^{1/6}\lambda_1 \bigg)+\text{Cov}\bigg( \frac{\lambda_1\|u_1\|_1^2}{n},n^{1/6}\lambda_1 \bigg).
\end{align*}
On the one hand, we have
\begin{align*}
\text{Cov}\bigg( \frac{\lambda_1\|u_1\|_1^2}{n},n^{1/6}\lambda_1 \bigg)&=n^{-5/6}(\E  \|u_1\|_1^2\lambda_1^2-\E \lambda_1\|u_1\|_1^2\E \lambda_1)\\
&=n^{-5/6}\E  \|u_1\|_1^2 \cdot \text{Var}(\lambda_1)\\
&\le n^{1/6}\E  \|u_1\|_1^2 \cdot \E (\lambda_1/\sqrt{n}-2)^2.
\end{align*}
Lemma 5 and the almost sure bound $\lambda_1= (2+o(1))\sqrt{n}$ given by \cite{bai1988necessary} imply
\beq \label{var.lambda1}
\E (\lambda_1/\sqrt{n}-2)^2=O(n^{-4/3+\epsilon}).
\eeq
Since $\|u_1\|_2=1$, we also have
\[
\frac{1}{n}\E \|u_1\|_1^2\le 1.
\]
Combining these results, we have
\beq \label{cov1}
\text{Cov}\bigg( \frac{\lambda_1\|u_1\|_1^2}{n},n^{1/6}\lambda_1 \bigg)=O(n^{-1/6+\epsilon}).
\eeq
On the other hand, note that
\[
A_n/n=\Omega_0+n^{-1/2}\sum_{i=2}^n(\lambda_i/n^{1/2}-\gamma_i)(\text{sgn}(u_1^\top)u_i)^2.
\]
Then
\begin{align*}
\text{Cov}\bigg( \frac{A_n}{n},n^{1/6}\lambda_1 \bigg)&=\text{Cov}\big( \Omega_0,n^{1/6}\lambda_1 \big)+\text{Cov}\bigg( n^{-1/2}\sum_{i=2}^n(\lambda_i/n^{1/2}-\gamma_i)(\text{sgn}(u_1^\top)u_i)^2,n^{1/6}\lambda_1 \bigg)\\
&=n^{-1/2}\sum_{i=2}^n\text{Cov}\bigg( (\lambda_i/n^{1/2})(\text{sgn}(u_1^\top)u_i)^2,n^{1/6}\lambda_1 \bigg),
\end{align*}
as a result of the independence between the eigenvalues and the eigenvectors of GOE. Now since, for any $i\ge 2$,
\begin{align*}
\text{Cov}\bigg( (\lambda_i/n^{1/2})(\text{sgn}(u_1^\top)u_i)^2,n^{1/6}\lambda_1 \bigg)&=n^{-1/3}\E \lambda_i(\text{sgn}(u_1^\top)u_i)^2\lambda_1-n^{-1/3}\E  \lambda_i(\text{sgn}(u_1^\top)u_i)^2\E \lambda_1\\
&=n^{-1/3}\E (\text{sgn}(u_1^\top)u_i)^2 \cdot \text{Cov}(\lambda_i,\lambda_1),
\end{align*}
where by Cauchy-Schwartz inequality
\[
\text{Cov}(\lambda_i,\lambda_1)\le \sqrt{\text{Var}(\lambda_i)\Var(\lambda_1)}.
\]
In particular, $\Var(\lambda_1)$ can be bounded using (\ref{var.lambda1}), and $\text{Var}(\lambda_i)$ for $i\ge 2$ has the upper bound
\[
\text{Var}(\lambda_i)=\text{Var}(\lambda_i-\sqrt{n}\gamma_i)\le \E (\lambda_i-\sqrt{n}\gamma_i)^2\le n\E(\lambda_i/\sqrt{n}-\gamma_i)^2=O(n^{-1/3+\epsilon}),
\]
where the last inequality follows from Lemma 5 and  (\ref{var.lambda1}). Moreover, using Lemma 1 and Lemma 6, we have $\E (\text{sgn}(u_1^\top)u_i)^2=O(1)$. Hence
\[
\text{Cov}\bigg( (\lambda_i/n^{1/2})(\text{sgn}(u_1^\top)u_i)^2,n^{1/6}\lambda_1 \bigg)=O(n^{-2/3+\epsilon}),
\]
and therefore
\beq \label{cov2}
\text{Cov}\bigg( \frac{A_n}{n},n^{1/6}\lambda_1 \bigg)=O(n^{-1/6+\epsilon}).
\eeq
The proof is complete by combining (\ref{cov1}) and (\ref{cov2}).

\section{Proof of Theorem 3}

Let $u_i$ and $\lambda_i(W)$ be the $i$-th eigenvector and eigenvalue of $W$. By definition, we have
\begin{align} \label{Qn.eq}
Q/n-2\|u_1\|_1^2/\sqrt{n}&=\textup{sgn}({u}_{1}^\top)W \textup{ sgn}({u}_{1})/n-2\|u_1\|_1^2/\sqrt{n}\nonumber\\ &={\lambda}_1(W)\{\textup{sgn}({u}_{1}^\top){u}_1\}^2/n-2\|u_1\|_1^2/\sqrt{n}+ \sum_{i=2}^n{\lambda}_i(W)\{\textup{sgn}({u}_{1}^\top){u}_i\}^2/n \nonumber \\
&=(\lambda_1(W)-2\sqrt{n})\|{u}_1\|_1^2/n+ \sum_{i=2}^n\lambda_i(W)\{\textup{sgn}({u}_{1}^\top){u}_i\}^2/n.
\end{align}
About the first term, by Weyl's perturbation inequality (Corollary III.2.6 of \cite{bhatia2013matrix}) and Bai-Yin's law \citep{bai1988necessary}, we have, for any $i=1,...,n,$
\[
|\lambda_i(W)-\lambda_i(\Theta)|\le \|Z\| \le C\sqrt{n},
\]
almost surely for some constant $C>0$. In other words,
\[
\lambda_1(W)\ge \lambda_1(\Theta)-C\sqrt{n}\gtrsim \sqrt{n}
\]
for some sufficiently large $C_0>C$. Now since $\|x\|_1\ge \sqrt{n}\|x\|_2$, we have
\[
\|{u}_1\|_1^2/n\ge 1,
\]
so that $(\lambda_1(W)-2\sqrt{n})\|{u}_1\|_1^2/n\gtrsim \sqrt{n}$ for sufficiently large $C_0$. Therefore, it suffices to show that $\sum_{i=2}^n\lambda_i(W)\{\textup{sgn}({u}_{1}^\top){u}_i\}^2/n\gtrsim-\sqrt{n}$. To see this, by Corollary III.2.2 of \cite{bhatia2013matrix}, we have
\[
\lambda_i(W)\ge \lambda_i(Z)+\lambda_n(\Theta).
\]
As a result,
\[
\sum_{i=2}^n\lambda_i(W)\{\textup{sgn}({u}_{1}^\top){u}_i\}^2/n\ge \sum_{i=2}^n(\lambda_i(Z)+\lambda_n(\Theta))\{\textup{sgn}({u}_{1}^\top){u}_i\}^2/n.
\]
Since $u_i$'s are orthonormal vectors, the vector $\tau=(\text{sgn}(u_1)^\top u_2/\sqrt{n},...,\text{sgn}(u_1)^\top u_n/\sqrt{n})$ satisfies $\|\tau\|_2^2\le 1$. Therefore, it holds that
\[
\sum_{i=2}^n(\lambda_i(Z)+\lambda_n(\Theta))\{\textup{sgn}({u}_{1}^\top){u}_i\}^2/n\ge \lambda_n(Z)+\lambda_n(\Theta)\gtrsim -\sqrt{n},
\]
almost surely. This implies that the test statistic $Q/n-2\|u_1\|_1^2/\sqrt{n}\to \infty$ almost surely as $n\to \infty$, which completes the proof.

\section{Supplementary Figures and Tables}

\subsection{More Simulations about Universality}

In this section, we provide more simulation results on the universality of our theoretical limit distribution considered in Section 3.2 of the main paper. Table \ref{table:t3}-\ref{table:t6} provide the empirical tail probabilities of the standardized statistic at different quantiles of $N(0,2(1-2/\pi)^2)$ under the following non-GOE settings: (i) the symmetric random matrices with heavy-tailed, nonsymmetric distributions such as the exponential distribution Exp(1); (ii) the adjacency matrix of sparse Erd{\H{o}}s-R\'enyi random graph \citep{erdHos2012spectral,erdHos2013spectral} with $p=n^{-1/4}$; and (iii) the sample correlation matrix $R_n$ of $N$ i.i.d. observations from $N(0,I_n)$ with $N=n^{5/2}$. In case (i) and (ii), the entries of the random matrices are normalized to match the first two moments of GOE. In case (iii), the modularity is calculated from the normalized matrix $\sqrt{N}(R_n-I_n)$. 

\begin{table}[h!]
	\centering
	\caption{Empirical tail probabilities at $N(0,2(1-2/\pi)^2)$ quantiles: the case of heavy-tailed distribution Exp(1)}
	\begin{tabular}{c|cccccc}
		\hline
		n & 50 & 100 & 500 & 1000 & 2000  & 5000\\
		\hline
		$\alpha=0\cdot$01 & $0\cdot$0868 & $0\cdot$0731 &  $0\cdot$0181 & $0\cdot$0124& $0\cdot$0098 & $0\cdot$0079 \\
		
		$\alpha=0\cdot$05 & $0\cdot$1531 & $0\cdot$1432 &  $0\cdot$0620 & $0\cdot$0498 & $0\cdot$0417&  $0\cdot$0379\\
		
		$\alpha=0\cdot$25 & $0\cdot$3089 & $0\cdot$3186 &  $0\cdot$2314& $0\cdot$2079 & $0\cdot$1963 &  $0\cdot$1912\\
		
		$\alpha=0\cdot$50 & $0\cdot$4585 & $0\cdot$4879 &  $0\cdot$4327 & $0\cdot$4116 & $0\cdot$4043 &  $0\cdot$4030\\
		
		$\alpha=0\cdot$75 & $0\cdot$6214 & $0\cdot$6658 & $0\cdot$6557 & $0\cdot$6457 & $0\cdot$6415 & $0\cdot$6468\\
		
		$\alpha=0\cdot$95 & $0\cdot$8242 & $0\cdot$8695 &  $0\cdot$8874 & $0\cdot$8913 & $0\cdot$8949 & $0\cdot$9005\\
		
		$\alpha=0\cdot$99 & $0\cdot$9166 & $0\cdot$9477 & $0\cdot$9655& $0\cdot$9668 & $0\cdot$9694 &  $0\cdot$9724\\
		\hline
	\end{tabular}
	\label{table:t3}
\end{table}

\begin{table}[h!]
	\centering
	\caption{Empirical tail probabilities at $N(0,2(1-2/\pi)^2)$ quantiles: the case of Erd\"os-Renyi random graph ($p=n^{-1/4}$)}
	\begin{tabular}{c|cccccc}
		\hline
		n & 50 & 100 & 500 & 1000 & 2000  & 5000\\
		\hline
		$\alpha=0\cdot$01 & $0\cdot$0005 & $0\cdot$0016 &  $0\cdot$0048 & $0\cdot$0063& $0\cdot$0072 & $0\cdot$0080 \\
		
		$\alpha=0\cdot$05 & $0\cdot$0029 & $0\cdot$0071 &  $0\cdot$0222 & $0\cdot$0283 & $0\cdot$0333&  $0\cdot$0380 \\
		
		$\alpha=0\cdot$25 & $0\cdot$0236 & $0\cdot$0460 &  $0\cdot$1193& $0\cdot$1465 & $0\cdot$1687 &  $0\cdot$1887 \\
		
		$\alpha=0\cdot$50 & $0\cdot$0764 & $0\cdot$1309 &  $0\cdot$2743 & $0\cdot$3273 & $0\cdot$3629 &  $0\cdot$4025\\
		
		$\alpha=0\cdot$75 & $0\cdot$1959 & $0\cdot$2921 & $0\cdot$4948 & $0\cdot$5617 & $0\cdot$6060 & $0\cdot$6475\\
		
		$\alpha=0\cdot$95 & $0\cdot$4923 & $0\cdot$6118 &  $0\cdot$8015 & $0\cdot$8472 & $0\cdot$8764 & $0\cdot$9017 \\
		
		$\alpha=0\cdot$99 & $0\cdot$7175 & $0\cdot$8096 & $0\cdot$9271 & $0\cdot$9492 & $0\cdot$9621 &  $0\cdot$9736 \\
		\hline
	\end{tabular}
	\label{table:t5}
\end{table} 

\begin{table}[h!]
	\centering
	\caption{Empirical tail probabilities at $N(0,2(1-2/\pi)^2)$ quantiles: the case of sample correlation matrix ($N=n^{5/2}$)}
	\begin{tabular}{c|cccccc}
		\hline
		n & 20 & 75 & 50 & 100 & 150 & 200  \\
		\hline
		$\alpha=0\cdot$01 & $0\cdot$0028 & $0\cdot$0063 &  $0\cdot$0058 & $0\cdot$0067& $0\cdot$0065 & $0\cdot$0073 \\
		
		$\alpha=0\cdot$05 & $0\cdot$0102 & $0\cdot$0201 &  $0\cdot$0214 & $0\cdot$0230 & $0\cdot$0222&  $0\cdot$0282 \\
		
		$\alpha=0\cdot$25 & $0\cdot$0561 & $0\cdot$0902 &  $0\cdot$1066 & $0\cdot$1096 & $0\cdot$1252 &  $0\cdot$1344 \\
		
		$\alpha=0\cdot$50 & $0\cdot$1322 & $0\cdot$2035 &  $0\cdot$2399 & $0\cdot$2536 & $0\cdot$2775 &  $0\cdot$2917 \\
		
		$\alpha=0\cdot$75 & $0\cdot$2646 & $0\cdot$3880 & $0\cdot$4343 & $0\cdot$4463 & $0\cdot$4955 & $0\cdot$5060\\
		
		$\alpha=0\cdot$95 & $0\cdot$5559 & $0\cdot$6913 &  $0\cdot$7397 & $0\cdot$7507 & $0\cdot$7926 & $0\cdot$8098 \\
		
		$\alpha=0\cdot$99 & $0\cdot$7617 & $0\cdot$8610 & $0\cdot$8835 & $0\cdot$9011 & $0\cdot$9188 &  $0\cdot$9287 \\
		\hline
	\end{tabular}
	\label{table:t6}
\end{table} 

\subsection{Details about the Two Alternative Methods}

In this section, we provide more details about the two alternative tests considered in Section 3.3 of the main paper, namely, the \emph{ Largest Eigenvalue Test} based on $\lambda_1(W)$ and its Tracy-Widom limiting distribution \citep{johnstone2012fast}, and the \emph{Entrywise Maximum Test} based on the entrywise maxima $\max_{1\le i\ne j\le n}|\{\text{cov}(W)\}_{ij}|$ of the covariance matrix and its Gumbel limiting distribution \citep{jiang2004asymptotic,hu2020using}. Specifically, under the GOE null, we know that $\lambda_1(W)$ converges weakly to a standard Tracy-Widom random variable, while $T_n=\max_{1\le i\ne j\le n}|\{\text{cov}(W)\}_{ij}|$ has the following limiting distribution
\[
P(nT_n^2-4\log n+\log\log n \le y)\to \exp(-K\exp(-y/2)),\qquad \text{as $n\to\infty$,}
\]
where $K=\surd(8\pi)$. Both of the tests reject the null hypothesis whenever the test statistic exceeds the top $(1-\alpha)$ percentile of the limiting distribution. Table \ref{table:t9} provides the empirical type I errors of these two tests at the level $\alpha=0.05$, showing the asymptotic validity of these tests. The empirical type I errors are calculated based on 100,000 rounds of simulations.

\begin{table}[h!]
	\centering
	\caption{Empirical type I errors of the two alternatives methods at level $\alpha=0.05$}
	\begin{tabular}{c|cccccc}
		\hline
		$n$ & 50  &  100 & 200 & 400 & 600 & 800 \\
		\hline
		Largest Eigenvalue Test & 0.0389 & 0.0399 &   0.0397 & 0.0456 & 0.0456 & 0.0440 \\
		
		Entrywise Maximum Test & 0.1720 & 0.1233 &  0.0902 & 0.0700 & 0.0659 &  0.0642 \\
		\hline
	\end{tabular}
	\label{table:t9}
\end{table}

\subsection{Comparisons of Modularity and Normalized Modularity}

\begin{figure}
	\centering
	\includegraphics[angle=0,width=5in]{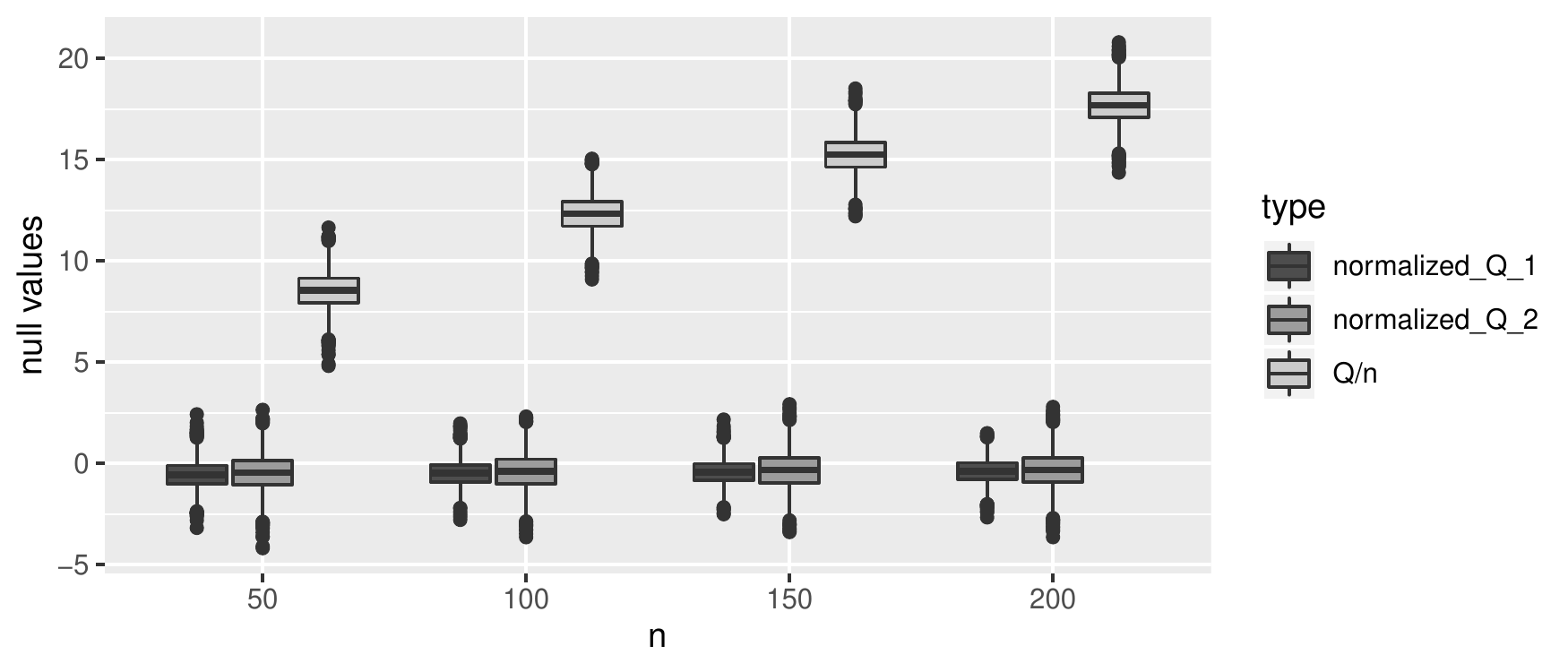}
	\includegraphics[angle=0,width=5in]{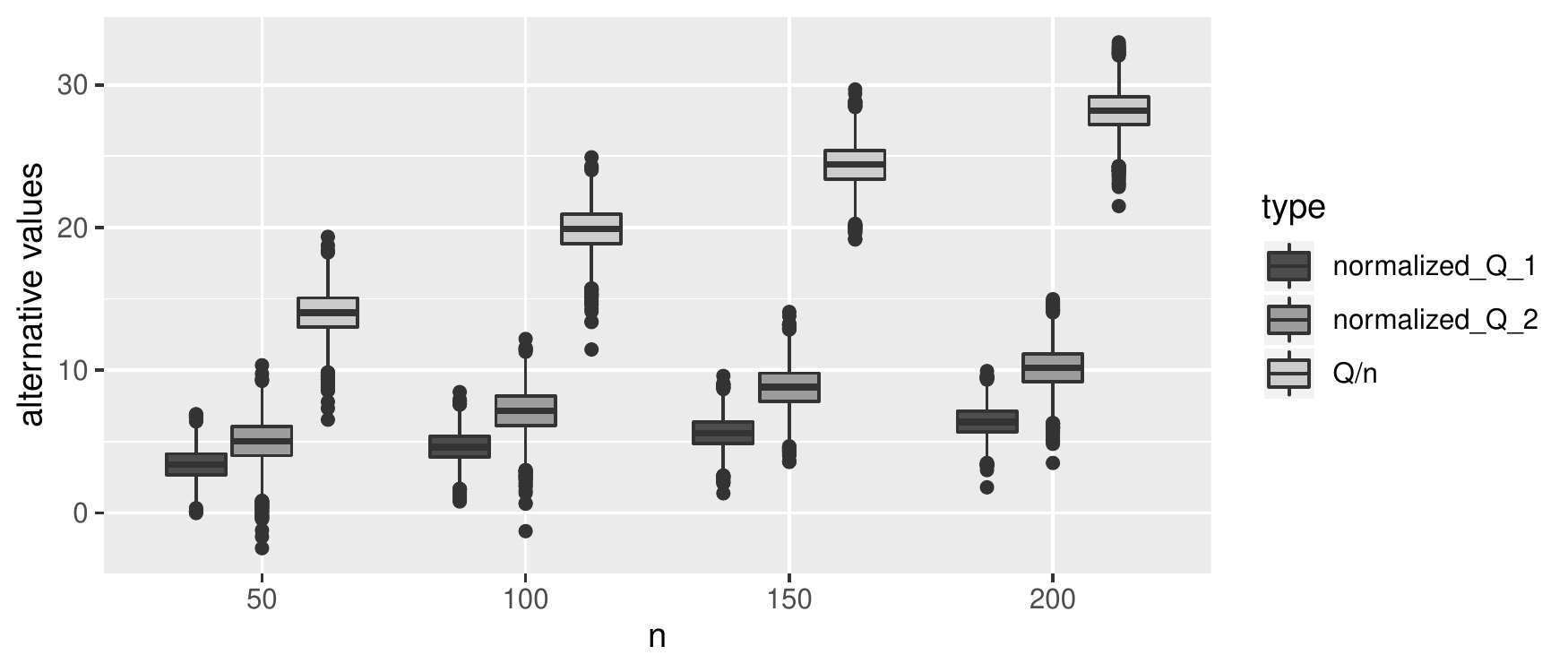}
	\includegraphics[angle=0,width=5in]{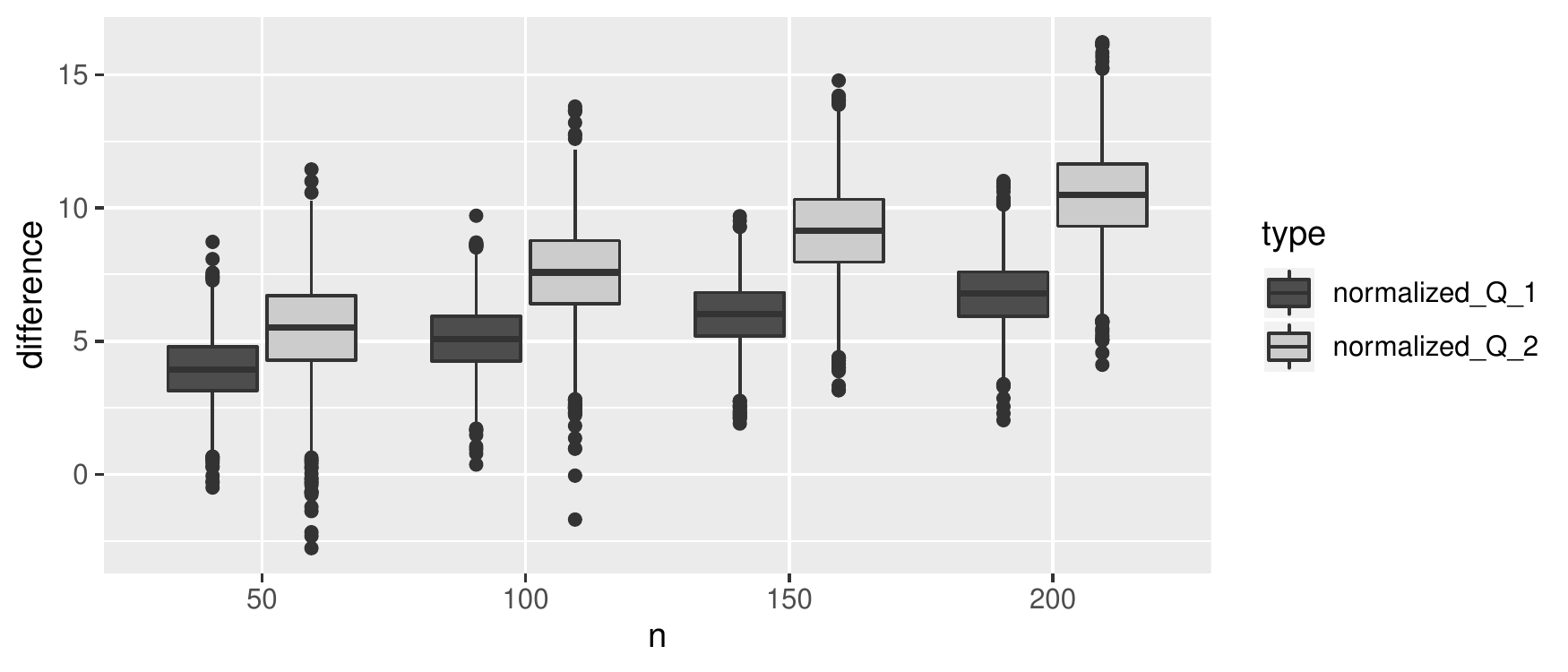}
	\caption{{Numerical comparisons of the modularities under the null and the alternative models.}  Top: boxplots of the modularities calculated from the standard GOE matrices (null model); Middle: boxplots of the modularities calculated from the deformed GOE matrices (alternative model); Bottom: boxplots of the differences of the two normalized modularities obtained from the null and the alternative models. All of the boxplots were produced from 3,000 rounds of simulations.}
	\label{Mod.fig}
\end{figure}

In Figure \ref{Mod.fig}, we compare numerically the scaled modularity $Q/n$ and the normalized modularities $n^{-1}(Q-2n^{1/2}\|u_1\|_1^2)$ (denoted as \verb|normalized_Q_1|) and $n^{-1}(Q-n^{3/2}4/\pi)$ (denoted as \verb|normalized_Q_2|) under the null and the alternative models. For each $n\in\{50,100,150,200\}$, we calculate these modularities from standard Wigner matrices (null), or the deformed GOE matrices (alternative) defined in Section 3.3 of our main paper with $\beta=2\sqrt{n}$, and produce the boxplots based on 3,000 rounds of simulations. From the top of Figure \ref{Mod.fig}, we find that under the null model, as $n$ grows, the magnitude of the modularity $Q/n$ increase, whereas empirical distributions of the normalized modularities remain roughly the same. Under the alternative model, in the middle of Figure \ref{Mod.fig}, all the three modularities underwent some mean shifts as $n$ increases, with relatively stable variances. This implies that, to construct a valid statistical test, or to better compare the evidences of network community structures, especially across studies with different sample sizes, one should use the normalized modularities instead of $Q$ or $Q/n$. In addition, we observe that under both null and alternative, \verb|normalized_Q_1| has slightly smaller variability than \verb|normalized_Q_2|. At the bottom of Figure \ref{Mod.fig}, we compare the differences of the two normalized modularities calculated from the null and alternative models. We find that the differences is more significant for \verb|normalized_Q_2| than those for  \verb|normalized_Q_1|, which suggests that a test based on \verb|normalized_Q_2| could be potentially more powerful under such alternatives. 

\label{lastpage}

\end{document}